\documentclass[a4paper,11pt]{article}
\pdfoutput=1
\usepackage{jheppub}
\usepackage[T1]{fontenc}

\makeatletter
\gdef\@fpheader{Prepared for submission to JHEP \hfill DIAS-STP-26-03}
\makeatother

\title{\boldmath Gravitational scalar production with a generic reheating scenario}

\author[a]{Francesco Costa}
\author[b,*]{and Jinsu Kim}
\note[*]{Corresponding author.}

\affiliation[a]{
	School of Theoretical Physics, Dublin Institute for Advanced Studies, \\
	10 Burlington Road, Dublin, D04 C932, Ireland
}
\affiliation[b]{
    School of Physics, Faculty of Basic Sciences, University of Shanghai for Science and Technology,\\
    Shanghai 200093, China
}

\emailAdd{fcosta@stp.dias.ie}
\emailAdd{kimjinsu@usst.edu.cn}

\abstract{
Gravitational production of decoupled scalars during inflationary and post-inflationary phases is efficient and can lead to over-production. We study this production with various reheating scenarios such as a generic power-law inflaton potential $V_{\rm inf}\propto \phi^k$ as well as a multi-stage reheating scenario. We derive constraints on the scalar self-interaction coupling $\lambda_s$, the mass $m_s$, and coefficients of quantum gravity-induced operators. We find that the constraints depend sensitively on the reheating dynamics. Our analysis demonstrates that universal gravity effects do not necessarily spoil the predictivity of non-thermal dark matter scenarios with $k < 4$ and low reheating temperatures, as an extended reheating phase dilutes gravitationally-produced relics. For $k > 4$, on the other hand, the relic abundance is enhanced during the reheating phase, leading to stringent constraints on the scalar. In multi-stage reheating, we show that the enhancement/dilution effect of subsequent reheating phases factorises.
}

\begin{document}
\maketitle
\flushbottom

\section{Introduction}
\label{sec:intro}
Whilst the Standard Model (SM) provides a remarkably successful description of our Universe, it is well established that the SM is incomplete especially when confronted with cosmological observations. Measurements of the Cosmic Microwave Background (CMB) as well as large-scale structure consistently point to the existence of non-baryonic dark matter (DM), which accounts for approximately 26\% of the present energy density of the Universe \cite{Planck:2018vyg}. Despite decades of experimental and theoretical effort, the microscopic nature of DM remains unknown, with no confirmed non-gravitational interactions with the SM observed so far; see, {\it e.g.}, Ref.~\cite{Bertone:2004pz} for a comprehensive review.

For many years, Weakly Interacting Massive Particles (WIMPs), produced via the thermal freeze-out mechanism, have represented the dominant paradigm for the DM candidates \cite{Kolb:1990vq,Arcadi:2017kky}. However, the absence of experimental evidence for such candidates in direct/indirect detection probes and collider searches has motivated increasing interest in non-thermal production mechanisms. Amongst these, the freeze-in mechanism has emerged as a well-motivated alternative \cite{McDonald:2001vt,Hall:2009bx}. The freeze-in mechanism assumes an initially negligible abundance and extremely feeble interactions with the thermal bath, such that DM is gradually produced through rare SM decays or annihilations without ever reaching equilibrium. In contrast to freeze-out, the final relic abundance increases with the interaction strength, and the DM candidate is commonly referred to as Feebly Interacting Massive Particles (FIMPs).

A fundamental challenge for non-thermal DM scenarios arises from gravitational particle production. In a time-dependent spacetime background, particles can be generated purely through gravitational effects, even in the absence of apparent couplings \cite{Mamaev:1976zb,Ford:1986sy}. Notably, during inflation, which offers a compelling explanation for the observed homogeneity, isotropy, and flatness of the Universe, as well as the nearly scale-invariant spectrum of primordial curvature perturbations measured in the CMB \cite{Guth:1980zm,Linde:1981mu,Mukhanov:1981xt,Albrecht:1982wi}, light spectator fields generically acquire superhorizon fluctuations that translate into non-negligible post-inflationary abundance \cite{Starobinsky:1986fx,Starobinsky:1994bd}. Additional production may occur during reheating after the end of inflation \cite{Abbott:1982hn,Kofman:1997yn,Greene:1997fu,Chung:1998rq,Ichikawa:2008ne,Ema:2016hlw,Cembranos:2019qlm,Garcia:2020wiy,Mambrini:2021zpp,Lebedev:2022ljz,Lebedev:2022cic,Feiteira:2025rpe}. In some cases, such mechanisms can account for the observed DM abundance on their own \cite{Peebles:1999fz,Nurmi:2015ema,Markkanen:2018gcw}. More generally, however, gravitational production poses a challenge to the basic assumption underlying the freeze-in mechanism, namely the negligible initial DM population. Readers may refer to, {\it e.g.}, Refs.~\cite{Ford:2021syk,Kolb:2023ydq} for a review on gravitational particle production.

Importantly, even in scenarios where the dark sector is completely decoupled from the visible sector of the SM at the renormalisable level, gravitational effects cannot be consistently neglected. At high energy scales such as our early Universe, we expect quantum gravity to play a non-negligible role. One may parametrise the effect of quantum gravity in an effective field theory way through Planck-suppressed operators \cite{Donoghue:1994dn,Burgess:2003jk,Lebedev:2022ljz}. Such operators generically couple the inflaton to spectator fields and therefore provide an irreducible source of DM production. The production associated with the quantum gravity-induced operators is typically very efficient, and to avoid over-production often requires extremely small Wilson coefficients \cite{Lebedev:2022cic}. The viability of freeze-in DM models thus depends crucially on the relic abundance produced via gravity effects during the early epochs and on the detailed physics of reheating, as we illustrate in this work.

Scenarios with a low reheating temperature have attracted renewed attention in this context. A sufficiently low reheating temperature can efficiently dilute gravitationally-produced relics, thereby restoring the consistency of freeze-in and allowing viable DM production for interaction strengths far larger than in the conventional high-temperature picture \cite{Cosme:2023xpa,Cosme:2024ndc,Costa:2024ugy}. Once largely overlooked, this regime has recently been explored in a broad range of models, including Higgs-portal and more general dark sector constructions; see, {\it e.g.}, Refs.~\cite{Gan:2023jbs,Arcadi:2024wwg,Silva-Malpartida:2023yks,Henrich:2024rux,Belanger:2024yoj,Bernal:2024ndy,Silva-Malpartida:2024emu,Bernal:2024yhu,Lebedev:2024mbj,Arcadi:2024obp,Koivunen:2024vhr,Boddy:2024vgt,Khan:2024biq,Lee:2024wes,Barman:2024nhr,Barman:2024tjt,Borah:2025ema,Bernal:2025osg,Khan:2025keb,Arias:2025tvd,Arias:2025nub,Bernal:2025qkj,Khan:2025kuh}. In this low-reheating regime, the DM--SM coupling is allowed to be as large as $\mathcal{O}(1)$ whilst avoiding thermalisation, making these scenarios experimentally interesting and observationally testable.

In this work, we provide a comprehensive analysis of gravitational scalar production, considering a general reheating scenario. We systematically examine both production during inflation and production during reheating, paying particular attention to how different reheating dynamics affect the final relic abundance of the scalar. Our analysis encompasses a range of reheating scenarios, from the simple instantaneous reheating to realistic and generic multi-stage reheating. We derive general constraints on scalar self-interaction coupling and on quantum gravity-induced couplings as a function of the reheating temperature and the form of the inflaton potential.

The paper is organised as follows. In Sec.~\ref{sec:prodinf}, we study the production of a scalar during inflation and track the evolution of the resulting condensate across various reheating scenarios, including instantaneous, single-stage, and multi-stage reheating. In Sec.~\ref{sec:prodreh}, we analyse gravitational particle production during reheating induced by higher-dimensional, quantum gravity-induced inflaton--scalar operators, again allowing for general reheating dynamics. We summarise our findings and present our conclusions in Sec.~\ref{sec:conc}, where we also briefly discuss the impact of other possible interactions between the scalar and the SM or the inflaton. Several technical details and complementary results are relegated to the appendices.

\section{Production during inflation}
\label{sec:prodinf}
We first consider the production of a scalar $s$ during inflation. The potential for the scalar field $s$ is given by
\begin{align}
	V(s) = \frac{1}{2} m_s^2 s^2 + \frac{1}{4} \lambda_s s^4
	\,,
\end{align}
where the scalar mass is taken to be much smaller than the Hubble parameter during inflation. As our focus is on feebly-interacting particles, we further assume that the self-interaction coupling is small, {\it i.e.}, $\lambda_s \ll 1$, but not zero.\footnote{
	The free scalar field case, $\lambda_s = 0$, is qualitatively different from the case under consideration, namely $\lambda_s \ll 1$ (with $\lambda_s \langle s^2 \rangle_{\rm end} \gg m_s^2$, see also footnote~\ref{footnote:lambdasneq0-2}). Thus, the analysis presented here cannot be used to infer any constraint on such a scenario. For instance, the Starobinsky--Yokoyama equilibrium variance would be given, in the free scalar case, by $\langle s^2 \rangle_{\rm end} = 3 H_{\rm end}^4 / (8 \pi^2 m_s^2)$ \cite{Starobinsky:1994bd,Tenkanen:2019aij,Feiteira:2025rpe} instead of Eq.~\eqref{eqn:fieldvar_SY}. Readers may refer to Ref.~\cite{Feiteira:2025rpe} for the corresponding constraints in the free scalar case.
} 
When the scalar mass is lighter than the inflationary Hubble scale, quantum fluctuations of the field get continuously generated. Following the standard Starobinsky--Yokoyama approach to inflation \cite{Starobinsky:1994bd}, one can see that the stochastic noise from horizon-crossing modes builds up a non-zero variance in the field value. At the end of inflation, the variance $\langle s^2 \rangle_{\rm end}$ can be expressed as
\begin{align}
	\langle s^2 \rangle_{\rm end} = \alpha_{\rm SY} H_{\rm end}^2
	\,.\label{eqn:fieldvar_SY}
\end{align}
The coefficient $\alpha_{\rm SY}$ encodes whether the duration of inflation is long enough for the probability distribution function of $s$ to reach its equilibrium. When equilibrium is reached, one obtains \cite{Starobinsky:1994bd}
\begin{align}
    \alpha_{\rm SY} =
    \frac{\Gamma(3/4)}
    {\Gamma(1/4)}\sqrt{\frac{3}{2\pi^2}}\frac{1}{\sqrt{\lambda_s}}
    \simeq
    \frac{0.1}{\sqrt{\lambda_s}}
	\,.
	\label{eqn:alphaSYeq}
\end{align}
On the other hand, when inflation is not long enough for equilibrium, one may take $\alpha_{\rm SY} \simeq 1$, which can be considered as the lower bound \cite{Felder:1999wt,Lebedev:2022cic,Feiteira:2025rpe}. The dynamics of the scalar produced in this way is studied in detail in Refs.~\cite{Markkanen:2018gcw,Lebedev:2022cic,Feiteira:2025rpe}. In the present paper, we generalise the analysis of these studies to arbitrary reheating scenarios. For completeness, we start from the standard, simple reheating scenario, presenting the full derivation.

After the end of inflation, the Hubble parameter decreases, and at some point, it becomes comparable to the effective mass of $s$,
\begin{align}
	H^2 \simeq m_{\rm eff}^2 = m_s^2 + 3 \lambda_s \langle s^2 \rangle_{\rm end} = m_s^2 + 3 \alpha_{\rm SY} \lambda_s H_{\rm end}^2
	\,.
\end{align}
The condensate $s$ then starts to oscillate around its potential. If $m_s^2 \ll 3 \lambda_s \langle s^2 \rangle_{\rm end}$ at this point\footnote{
	The limit of $\lambda_s \to 0$, in which case the condition $m_s^2 \ll 3 \lambda_s \langle s^2 \rangle_{\rm end}$ cannot be met, is discussed in Ref.~\cite{Feiteira:2025rpe}; we do not consider such a free scalar in the present work. \label{footnote:lambdasneq0-2}
}, the condensate would oscillate around the quartic potential, and thus,
\begin{align}
	\bar{s} \propto \frac{1}{a}
	\,,
\end{align}
where $\bar{s} \equiv \sqrt{\langle s^2 \rangle}$. In other words, the scalar field $s$ behaves as radiation. We indicate the point at which the condensate $s$ starts to oscillate with the subscript ``osc''.

As time progresses, $\bar{s}$ decreases, and eventually, the bare mass term would dominate. The condensate $s$ then oscillates around its quadratic potential. We indicate the point at which this transition occurs with the subscript ``dust''. The transition happens when
\begin{align}
	\frac{1}{2} m_s^2 \bar{s}^2_{\rm dust} \simeq
	\frac{1}{4} \lambda_s \bar{s}_{\rm dust}^4
	\,,
\end{align}
and thus,
\begin{align}
	\bar{s}_{\rm dust} \simeq 
	\sqrt{\frac{2}{\lambda_s}} m_s
	\,.
\end{align}
From this point, the condensate would behave as non-relativistic dust, and thus, its energy density scales as
\begin{align}
	\rho(s) \propto \frac{1}{a^3}
	\,.
\end{align}
In the final dust stage, the number density is given by
\begin{align}
	n_s(a) = \frac{\rho(s)}{m_s} \simeq 
	\frac{m_s^3}{\lambda_s} \frac{a_{\rm dust}^3}{a^3}
	\,,
\end{align}
and the yield can be evaluated through
\begin{align}
	Y = \frac{n_s}{s_{\rm SM}}
	\,,
\end{align}
with the entropy density of the SM thermal bath at temperature $T$,
\begin{align}
	s_{\rm SM} = \frac{2\pi^2}{45} g_{*s} T^3
	\,,
\end{align}
where $g_{*s}$ is the effective entropy degrees of freedom.

The evaluation of the yield $Y$ requires information on the reheating dynamics. We consider five different scenarios: (1) the instantaneous reheating, (2) reheating with a matter-dominated phase, (3) reheating with a generic power-law inflaton potential, (4) two-stage reheating scenario, and (5) multi-stage reheating scenario.

\subsection{Instantaneous reheating}
\label{subsec:instreh}
\begin{figure}
	\centering
	\includegraphics[width=1\linewidth]{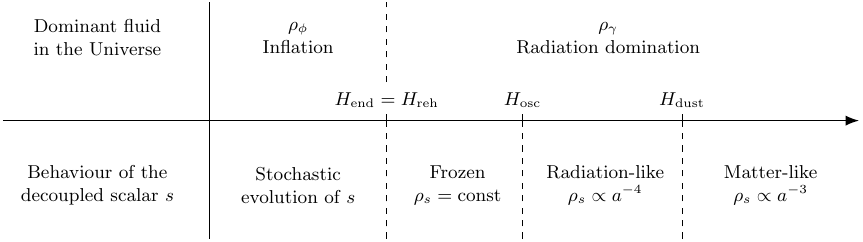}
	\caption{Schematic diagram for the evolution of the dominant energy component and a decoupled scalar. The instantaneous reheating scenario, where the energy of the inflaton gets immediately transferred to radiation at the end of inflation, is assumed here.}
	\label{fig:timeline-instreh}
\end{figure}
In the simplest scenario of instantaneous reheating, the energy of the inflaton gets immediately transferred to radiation at the end of inflation. Figure~\ref{fig:timeline-instreh} summarises the evolution of the scalar $s$ and the background in the instantaneous reheating case. As discussed earlier, the scalar $s$ first oscillates around its quartic potential, acting like radiation, as the Hubble parameter becomes lower than the effective mass of $s$. It then oscillates around its quadratic potential, acting like dust, when the bare mass term dominates the self-interaction term. It is worth noting that the scalar $s$ never dominates the energy density of the Universe at early times.

Once $s$ becomes non-relativistic, and in the absence of entropy injection, the yield $Y$ is conserved. We may thus evaluate the yield right at the onset of the dust period, {\it i.e.},
\begin{align}
	Y =
	\frac{n_{s,{\rm dust}}}{s_{\rm SM}(T_{\rm dust})}
	\,,
\end{align}
where $n_{s,{\rm dust}} \equiv n_s(a = a_{\rm dust}) \simeq m_s^3 / \lambda_s$. The temperature $T_{\rm dust}$ at the onset of the dust stage can be replaced by the reheating temperature $T_{\rm reh}$, utilising the entropy conservation. We then obtain
\begin{align}
	Y =
	\frac{m_s^3}{\lambda_s}
	\frac{(a_{\rm dust}/a_{\rm reh})^3}{s_{\rm SM}(T_{\rm reh})}
	=
	\frac{m_s^3}{\lambda_s}
	\frac{(a_{\rm dust}/a_{\rm end})^3}{s_{\rm SM}(T_{\rm reh})}
	\,,
\end{align}
where we have used $a_{\rm reh} = a_{\rm end}$ for the instantaneous reheating scenario. During the radiation-dominated epoch, the Hubble parameter scales as $H\propto a^{-2}$, and thus, we may write
\begin{align}
	\frac{a_{\rm osc}}{a_{\rm end}} =
	\sqrt{\frac{H_{\rm end}}{H_{\rm osc}}}
	\,,
\end{align}
where $H_{\rm osc} \simeq \sqrt{3 \lambda_s} \bar{s}_{\rm end}$. Furthermore, since $\bar{s} \propto a^{-1}$ during the regime where the scalar condensate $s$ oscillates around its quartic potential, we find
\begin{align}
	\frac{a_{\rm dust}}{a_{\rm osc}} =
	\frac{\bar{s}_{\rm osc}}{\bar{s}_{\rm dust}} \simeq
	\frac{\bar{s}_{\rm end}}{\bar{s}_{\rm dust}}
	\,,
\end{align}
where we have used the fact that the condensate $s$ stays frozen between the end of inflation and the start of the oscillation. Therefore,
\begin{align}
	\frac{a_{\rm dust}}{a_{\rm end}} \simeq
	\sqrt{\frac{H_{\rm end}}{H_{\rm osc}}}
	\frac{\bar{s}_{\rm end}}{\bar{s}_{\rm dust}} \simeq
	\frac{(\alpha_{\rm SY} \lambda_s / 3)^{1/4}}{\sqrt{2} m_s} H_{\rm end}
	\,.
\end{align}
In the instantaneous reheating scenario, the reheating temperature $T_{\rm reh}$ is also expressed in terms of the Hubble parameter at the end of inflation. Using the Friedmann equation, since $3H_{\rm end}^2 M_{\rm P}^2 = (\pi^2/30) g_* T_{\rm reh}^4$, where $g_*$ is the effective relativistic degrees of freedom for the energy density, we read
\begin{align}
\label{eq:T_reh_H_reh}
	T_{\rm reh} =
	\left(\frac{90}{\pi^2 g_*}\right)^{\frac{1}{4}} \sqrt{H_{\rm end}M_{\rm P}}
	\,.
\end{align}
We thus conclude that the yield of the scalar $s$ is given by
\begin{align}
	Y \simeq
	\frac{0.35 g_*^{3/4}}{g_{*s}}
	\frac{\alpha_{\rm SY}^{3/4}}{\lambda_s^{1/4}}
	\left(\frac{H_{\rm end}}{M_{\rm P}}\right)^{\frac{3}{2}}
	\,.
\end{align}

In the case where the probability distribution function of $s$ reaches its equilibrium, using Eq.~\eqref{eqn:alphaSYeq}, the yield becomes
\begin{align}
	Y \simeq
	\frac{0.077 g_{*}^{3/4}}{g_{*s}}
	\left(
	\frac{H_{\rm end}}{M_{\rm P}}
	\right)^{\frac{3}{2}}
	\lambda_s^{-\frac{5}{8}}
	\,.
	\label{eqn:syieldInstReh}
\end{align}
In order to avoid over-producing the scalar $s$, the abundance of $s$ must not exceed the DM abundance today, which translates to
\begin{align}
	Y \lesssim
	\frac{4.4\times 10^{-10} \, \text{GeV}}{m_s}
	\,.
	\label{eqn:abundance-constraint}
\end{align}
Together with the yield \eqref{eqn:syieldInstReh}, we obtain the following constraint:
\begin{align}
    m_s \lambda_s^{-\frac{5}{8}}
    \lesssim 
    3.0 \times 10^{-8}
    \left(
    \frac{g_{*s}}{g_{*}^{3/4}}
    \right)
    \left(
    \frac{M_{\rm P}}{H_{\rm end}}
    \right)^{\frac{3}{2}}
    \,{\rm GeV}
    \,,
    \label{eqn:inst-reh-constraint}
\end{align}
that matches the result of Ref.~\cite{Lebedev:2022cic}; see also Ref.~\cite{Markkanen:2018gcw}. The result for the non-equilibrium case is reported in Appendix~\ref{apdx:non-Eq}.

\begin{figure}
	\centering
	\includegraphics[width=1\linewidth]{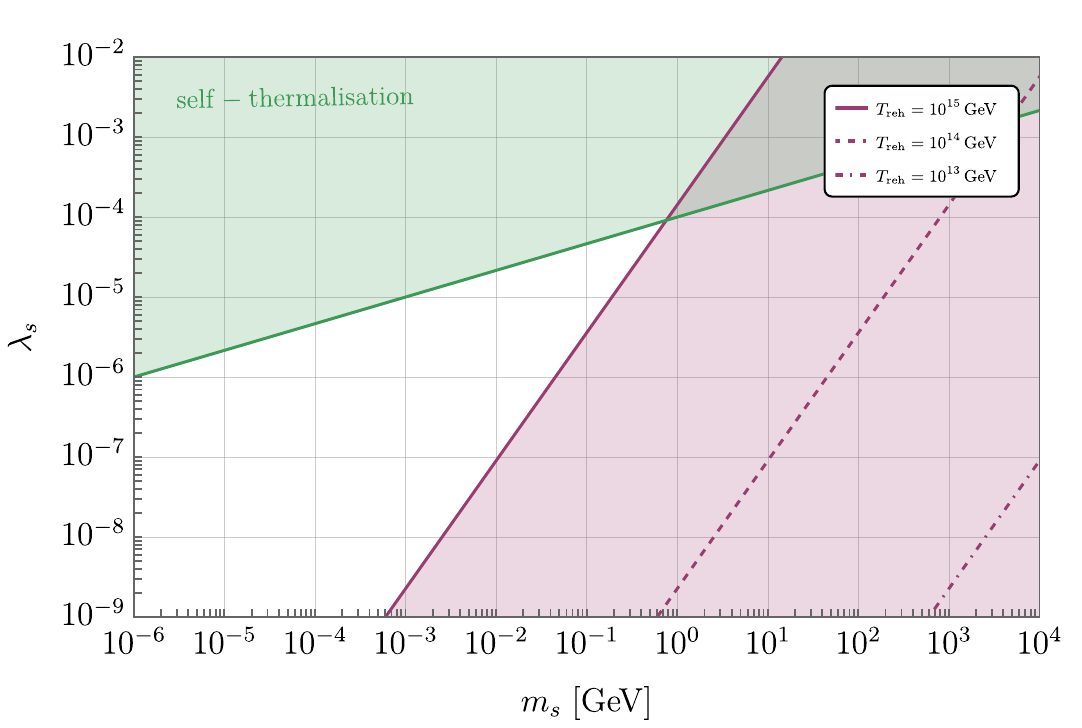}
	\caption{Constraints on the scalar self-interaction coupling $\lambda_s$ in terms of the scalar mass $m_s$ in the instantaneous reheating scenario. Together with the choice of $g_* = g_{*s} = 100$, three reheating temperatures are considered, namely $T_{\rm reh} = 10^{15}$ GeV (solid), $10^{14}$ GeV (dashed), and $10^{13}$ GeV (dot-dashed).	In terms of the Hubble parameter at the end of inflation, these values of the reheating temperature correspond to $H_{\rm end} = 1.3 \times 10^{12}$ GeV, $1.3 \times 10^{10}$ GeV, and $1.3 \times 10^{8}$ GeV, respectively. The self-thermalisation bound (green region) is adopted from Refs.~\cite{Arcadi:2019oxh,Lebedev:2022cic}.}
	\label{fig:Inst_reheating}
\end{figure}
In Fig.~\ref{fig:Inst_reheating}, the constraint \eqref{eqn:inst-reh-constraint} on the scalar self-interaction coupling $\lambda_s$ is presented in terms of the scalar mass $m_s$ in the case of instantaneous reheating. In this scenario, the Hubble parameter at the end of inflation $H_{\rm end}$ plays the same role as the reheating temperature $T_{\rm reh}$, as seen in Eq.~\eqref{eq:T_reh_H_reh}. Together with the choice of $g_* = g_{*s} = 100$, three different values of $T_{\rm reh}$ ($H_{\rm end}$) are considered in Fig.~\ref{fig:Inst_reheating}: $10^{15}$ GeV ($1.3 \times 10^{12}$ GeV), $10^{14}$ GeV ($1.3 \times 10^{10}$ GeV), and $10^{13}$ GeV ($1.3 \times 10^{8}$ GeV), depicted by solid, dashed, and dot-dashed lines, respectively.
As one may also see from Eq.~\eqref{eqn:syieldInstReh}, the yield is proportional to $H_{\rm end}$, and thus, lowering the inflationary scale significantly reduces the bound. We note that for $T_{\rm reh} = 10^{13}$ GeV, which corresponds to a low inflationary scale of $H_{\rm end} \simeq 10^8$ GeV, the mass range of GeV--TeV becomes available. This tendency of the inflationary scale dependence persists throughout the later cases as we shall see below.
The shaded green region represents the self-thermalisation bound, adopted from Refs.~\cite{Arcadi:2019oxh,Lebedev:2022cic}, which indicates the region where the scalar abundance is governed by the freeze-out mechanism within the dark sector due to self-thermalisation, breaking the feebly-interacting nature of the scalar. As this region contradicts with our setup, we exclude the parameter space allowing self-thermalisation.

\subsection{Matter domination during reheating}
\label{subsec:EMDreh}
Having discussed the simplest case of instantaneous reheating, we now move on to the case where the reheating phase after the end of inflation is described by a quadratic inflaton potential. We thus have a matter-dominated reheating epoch in this scenario.\footnote{
This scenario is equivalent to instantaneous reheating followed by an early matter-dominated epoch.
} We note that the evolution of the scalar $s$ remains unchanged. In this scenario, we assume that the matter-dominated reheating epoch lasts at least until after $s$ starts oscillating.\footnote{
When the matter-dominated reheating epoch terminates before the onset of $s$ oscillation, we obtain the same result as in the instantaneous reheating scenario, as we shall discuss in more detail in Sec.~\ref{subsec:PLreh}. \label{footnote:inst-reh-equiv}
}

During this matter-dominated reheating era, the Hubble parameter scales as $H\propto a^{-3/2}$. As before, we may compute the yield of $s$ as
\begin{align}
	Y =
	\frac{m_s^3}{\lambda_s}
	\frac{(a_{\rm dust}/a_{\rm reh})^3}{s_{\rm SM}(T_{\rm reh})}
	\,.
\end{align}
Unlike the instantaneous reheating scenario, however, $a_{\rm reh} \neq a_{\rm end}$ now, where the subscript ``reh'' refers to the point at which reheating ends. The end of the reheating phase happens after $a_{\rm osc}$, {\it i.e.}, $a_{\rm reh} > a_{\rm osc}$.\footref{footnote:inst-reh-equiv} The ratio $a_{\rm dust}/a_{\rm reh}$ can be separated into
\begin{align}
	\frac{a_{\rm dust}}{a_{\rm reh}}
	=
	\frac{a_{\rm osc}}{a_{\rm reh}}
	\frac{a_{\rm dust}}{a_{\rm osc}}
	\,.
\end{align}
During the quartic regime of the $s$ oscillation, $\bar s \propto a^{-1}$, and between $a_{\rm osc}$ and $a_{\rm reh}$, as we are in the matter-dominated reheating phase, $H\propto a^{-3/2}$. Thus, we find
\begin{align}
	\frac{a_{\rm dust}}{a_{\rm reh}}
	=
	\left(
	\frac{H_{\rm reh}}{H_{\rm osc}}
	\right)^{\frac{2}{3}}
	\frac{\bar{s}_{\rm end}\sqrt{\lambda_s/2}}{m_s}
	\,.
\end{align}

Using $H_{\rm osc} \simeq \sqrt{3\lambda_s} \bar{s}_{\rm end}$, $\bar{s}_{\rm end} = \sqrt{\alpha_{\rm SY}}H_{\rm end}$, and
\begin{align}
	T_{\rm reh} =
	\left(\frac{90}{\pi^2 g_*}\right)^{\frac{1}{4}} \sqrt{H_{\rm reh}M_{\rm P}}
	\,,
\end{align}
we obtain
\begin{align}
	Y \simeq 
	\frac{0.051 g_*^{3/4}}{g_{*s}}
	\frac{\sqrt{\alpha_{\rm SY}}}{\sqrt{\lambda_s}}
	\Delta_{\rm reh}^{-1}
	\left(
	\frac{H_{\rm end}}{M_{\rm P}}
	\right)^{\frac{3}{2}}
	\,,
	\label{eqn:EMDsyieldgen}
\end{align}
where, following Ref.~\cite{Lebedev:2022cic}, we have defined
\begin{align}
	\Delta_{\rm reh} \equiv \sqrt{\frac{H_{\rm end}}{H_{\rm reh}}}
	\,,
	\label{eqn:DeltaNR}
\end{align}
which quantifies the duration of the matter-dominated reheating epoch.

For the case of the equilibrium probability distribution function of $s$, Eq.~\eqref{eqn:EMDsyieldgen} reduces to
\begin{align}
	Y \simeq 
	\frac{0.019 g_*^{3/4}}{g_{*s}}
	\Delta_{\rm reh}^{-1}
	\left(
	\frac{H_{\rm end}}{M_{\rm P}}
	\right)^{\frac{3}{2}}
	\lambda_s^{-\frac{3}{4}}
	\,.
\end{align}
Imposing the abundance constraint \eqref{eqn:abundance-constraint} then gives the following bound on the mass and the self-interaction coupling of the scalar $s$:
\begin{align}
	m_s \lambda_s^{-\frac{3}{4}}
	\lesssim 
	2.4 \times 10^{-8}
	\left(
	\frac{g_{*s}}{g_{*}^{3/4}}
	\right)
	\Delta_{\rm reh}
	\left(
	\frac{M_{\rm P}}{H_{\rm end}}
	\right)^{\frac{3}{2}}
	\,{\rm GeV}
	\,,
	\label{eqn:EMDsyieldEQ}
\end{align}
recovering the result of Ref.~\cite{Lebedev:2022cic}.
We notice that the bound gets relaxed by the factor of $\Delta_{\rm reh}$. The bound becomes weaker as the end of reheating happens later.
The result for the non-equilibrium case is reported in Appendix~\ref{apdx:non-Eq}.

\subsection{Power-law potential reheating}
\label{subsec:PLreh}
One may generalise the previous analysis of the matter-dominated reheating epoch involving the quadratic inflaton potential. Let us consider a reheating stage where the inflaton after the end of inflation oscillates around the potential minimum of a power-law potential $ V_{\rm inf}(\phi) \propto \phi^k$. During this reheating stage, the Hubble parameter would then evolve as
\begin{align}
	H(a) = 
	H_{\rm end} \left(
	\frac{a_{\rm end}}{a}
	\right)^{\frac{3k}{k+2}}
	\,.
\end{align}

We first examine the case where reheating ends during the dust regime of the scalar condensate $s$ when it oscillates around its quadratic potential, {\it i.e.}, $a_{\rm reh} > a_{\rm dust}$. At the end of reheating, the number density is given by
\begin{align}
	n_{s,{\rm reh}} =
	n_{s,{\rm dust}}\left(
	\frac{a_{\rm dust}}{a_{\rm reh}}
	\right)^3
	\,,
\end{align}
where
\begin{align}
	n_{s,{\rm dust}} =
	\frac{\rho_{\rm dust}}{m_s} \simeq 
	\frac{m_s^3}{\lambda_s}
	\,.
\end{align}
Since $\bar{s} \propto a^{-1}$ during the quartic oscillation regime, we may express $a_{\rm dust}$ as
\begin{align}
	a_{\rm dust} = 
	a_{\rm osc} \frac{\bar{s}_{\rm osc}}{\bar{s}_{\rm dust}}
	\simeq
	a_{\rm osc} \frac{\bar{s}_{\rm end}}{\bar{s}_{\rm dust}}
	\,,
\end{align}
where we have used the fact that the condensate $s$ stays frozen between the end of inflation and the start of the oscillation.
Noting that
\begin{align}
	\frac{a_{\rm osc}}{a_{\rm reh}} =
	\left(
	\frac{H_{\rm reh}}{H_{\rm osc}}
	\right)^{\frac{k+2}{3k}}
	\,,
\end{align}
we obtain
\begin{align}
	n_{s,{\rm reh}} \simeq
	\frac{m_s^3}{\lambda_s}
	\left(
	\frac{H_{\rm reh}}{H_{\rm osc}}
	\right)^{\frac{k+2}{k}}
	\left(
	\frac{\bar{s}_{\rm end}}{\bar{s}_{\rm dust}}
	\right)^3
	\,.
\end{align}
As $\bar{s}_{\rm end} = \sqrt{\alpha_{\rm SY}} H_{\rm end}$, $\bar{s}_{\rm dust} \simeq \sqrt{2/\lambda_s} m_s$, and $H_{\rm osc} \simeq \sqrt{3 \lambda_s} \bar{s}_{\rm end}$, one can see that the yield, $Y = n_{s,{\rm reh}}/s_{\rm SM}(T_{\rm reh})$, is given by
\begin{align}
	Y \simeq
	\frac{0.089 g_{*}^{3/4}}{3^{1/k} g_{*s}}
	\alpha_{\rm SY}^{1-\frac{1}{k}}
	\lambda_s^{-\frac{1}{k}}
	\Delta_{\rm reh}^{1-\frac{4}{k}}
	\left(
	\frac{H_{\rm end}}{M_{\rm P}}
	\right)^{\frac{3}{2}}
	\,,
	\label{eqn:PLsyieldgen}
\end{align}
where $\Delta_{\rm reh} \equiv \sqrt{H_{\rm end}/H_{\rm reh}}$ represents the duration of the reheating epoch.

For the case of the equilibrium probability distribution function of $s$, the yield \eqref{eqn:PLsyieldgen} reduces to
\begin{align}
	Y \simeq 
	C_k \frac{g_*^{3/4}}{g_{*s}}
	\lambda_s^{-\frac{1}{2k}-\frac{1}{2}}
	\Delta_{\rm reh}^{1-\frac{4}{k}}
	\left(
	\frac{H_{\rm end}}{M_{\rm P}}
	\right)^{\frac{3}{2}}
	\,,
	\label{eqn:PLsyieldEQ}
\end{align}
where
\begin{align}
	C_k \equiv 
	2^{\frac{1}{2k}-\frac{15}{4}}
	3^{\frac{1}{2}-\frac{3}{2k}}
	5^{\frac{1}{4}}
	\pi^{\frac{1}{k}-\frac{3}{2}}
	\left[
	\frac{\Gamma(3/4)}{\Gamma(1/4)}
	\right]^{1-\frac{1}{k}}
	\,.
\end{align}
The abundance constraint \eqref{eqn:abundance-constraint} then translates to
\begin{align}
	m_s \lambda_s^{-\frac{1}{2k}-\frac{1}{2}}
	\lesssim 
	\frac{4.4\times 10^{-10}}{C_k}
	\left(
	\frac{g_{*s}}{g_*^{3/4}}
	\right)
	\Delta_{\rm reh}^{\frac{4}{k}-1}
	\left(
	\frac{M_{\rm P}}{H_{\rm end}}
	\right)^{\frac{3}{2}}
	\, {\rm GeV}
	\,.
	\label{eqn:PLconstraintEQ}
\end{align}
One may easily see that the $k = 2$ case correctly recovers the result of the matter-dominated reheating scenario, Eq.~\eqref{eqn:EMDsyieldEQ}, discussed in Sec.~\ref{subsec:EMDreh}.
The result for the non-equilibrium case is reported in Appendix~\ref{apdx:non-Eq}.

\begin{figure}
	\centering
	\includegraphics[width=1\linewidth]{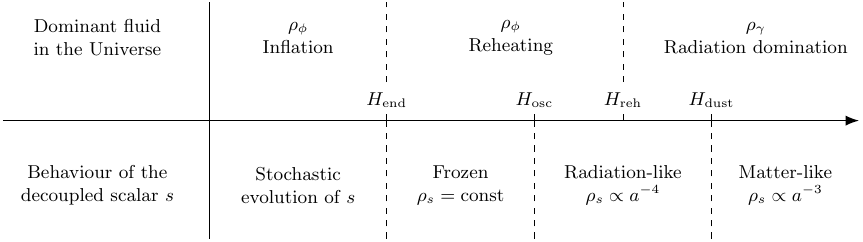}
	\caption{Schematic diagram for the evolution of the dominant energy component and the decoupled scalar $s$ across inflation, reheating, and radiation domination. Here, reheating is assumed to be completed before the onset of the dust regime of the scalar condensate.}
	\label{fig:timeline-reh}
\end{figure}

Next, we examine the $a_{\rm osc} < a_{\rm reh} < a_{\rm dust}$ case. The evolution of the dominant energy component and the scalar $s$ across inflation, reheating, and radiation-dominated phases is pictorially presented in Fig.~\ref{fig:timeline-reh}. In this case, the yield is evaluated at the transition point, $a_{\rm dust}$,
\begin{align}
    Y =
    \frac{n_{s,{\rm dust}}}{s_{\rm SM}(T_{\rm dust})}
    \,.
\end{align}
Using the scaling behaviour of the temperature, we may write
\begin{align}
    Y =
    \frac{n_{s,{\rm dust}}}{s_{\rm SM}(T_{\rm reh})}
    \left(
    \frac{a_{\rm dust}}{a_{\rm reh}}
    \right)^3
    =
    \frac{n_{s,{\rm reh}}}{s_{\rm SM}(T_{\rm reh})}
    \,,
\end{align}
which coincides with the yield we discussed above.

Finally, let us consider the $a_{\rm reh} < a_{\rm osc}$ case. The yield, evaluated at the transition point, $a_{\rm dust}$, is given by
\begin{align}
    Y =
    \frac{n_{s,{\rm dust}}}{s_{\rm SM}(T_{\rm dust})}
    = 
    \frac{n_{s,{\rm dust}}}{s_{\rm SM}(T_{\rm reh})}
    \left(
    \frac{a_{\rm dust}}{a_{\rm reh}}
    \right)^3
    \,.
\end{align}
Although the form is the same as the previous one, the scaling behaviour of $a_{\rm dust} / a_{\rm reh}$ differs. Since
\begin{align}
    a_{\rm dust} =
    a_{\rm osc}
    \frac{\bar{s}_{\rm osc}}{\bar{s}_{\rm dust}}
    \simeq
    a_{\rm osc}
    \frac{\bar{s}_{\rm end}}{\bar{s}_{\rm dust}}
    \,,
\end{align}
we have
\begin{align}
    \frac{a_{\rm dust}}{a_{\rm reh}} \simeq
    \frac{a_{\rm osc}}{a_{\rm reh}}
    \frac{\bar{s}_{\rm end}}{\bar{s}_{\rm dust}}
    \,.
\end{align}
When $a_{\rm reh} < a_{\rm osc}$, the ratio of the scale factors can be written as
\begin{align}
    \frac{a_{\rm osc}}{a_{\rm reh}} =
    \sqrt{\frac{H_{\rm reh}}{H_{\rm osc}}}
    \,,
\end{align}
which is the same as the instantaneous reheating scenario where the Universe goes straight into radiation domination. This is no coincidence; as reheating has already ended when the scalar condensate $s$ starts oscillating, the Universe appears to undergo the radiation-dominated epoch, and for the scalar $s$, it is no different from the instantaneous reheating case.

Thus, as the $a_{\rm reh} < a_{\rm osc}$ case recovers the instantaneous reheating case discussed in Sec.~\ref{subsec:instreh}, the only meaningful constraint one may extract from the power-law potential reheating scenario is the case where reheating terminates after the onset of the scalar $s$ oscillation, {\it i.e.}, $a_{\rm reh} > a_{\rm osc}$. For concreteness, we present explicit expressions of the constraint \eqref{eqn:PLconstraintEQ} for the $k = 2$, $k = 4$, and $k = 6$ cases:
\begin{align}
	m_s \lambda_s^{-\frac{3}{4}}
	&\lesssim 
	2.4\times 10^{-8}
	\left(
	\frac{g_{*s}}{g_*^{3/4}}
	\right)
	\Delta_{\rm reh}
	\left(
	\frac{M_{\rm P}}{H_{\rm end}}
	\right)^{\frac{3}{2}}
	\, {\rm GeV}
	&\quad \text{for } k = 2 \,,\\
	m_s \lambda_s^{-\frac{5}{8}}
	&\lesssim 
	3.0\times 10^{-8}
	\left(
	\frac{g_{*s}}{g_*^{3/4}}
	\right)
	\left(
	\frac{M_{\rm P}}{H_{\rm end}}
	\right)^{\frac{3}{2}}
	\, {\rm GeV}
	&\quad \text{for } k = 4 \,,\\
	m_s \lambda_s^{-\frac{7}{12}}
	&\lesssim 
	3.2\times 10^{-8}
	\left(
	\frac{g_{*s}}{g_*^{3/4}}
	\right)
	\Delta_{\rm reh}^{-\frac{1}{3}}
	\left(
	\frac{M_{\rm P}}{H_{\rm end}}
	\right)^{\frac{3}{2}}
	\, {\rm GeV}
	&\quad \text{for } k = 6 \,.
\end{align}
The corresponding expressions for the non-equilibrium case are summarised in Appendix~\ref{apdx:non-Eq}.

\begin{figure}
	\centering
	\includegraphics[width=0.9\linewidth]{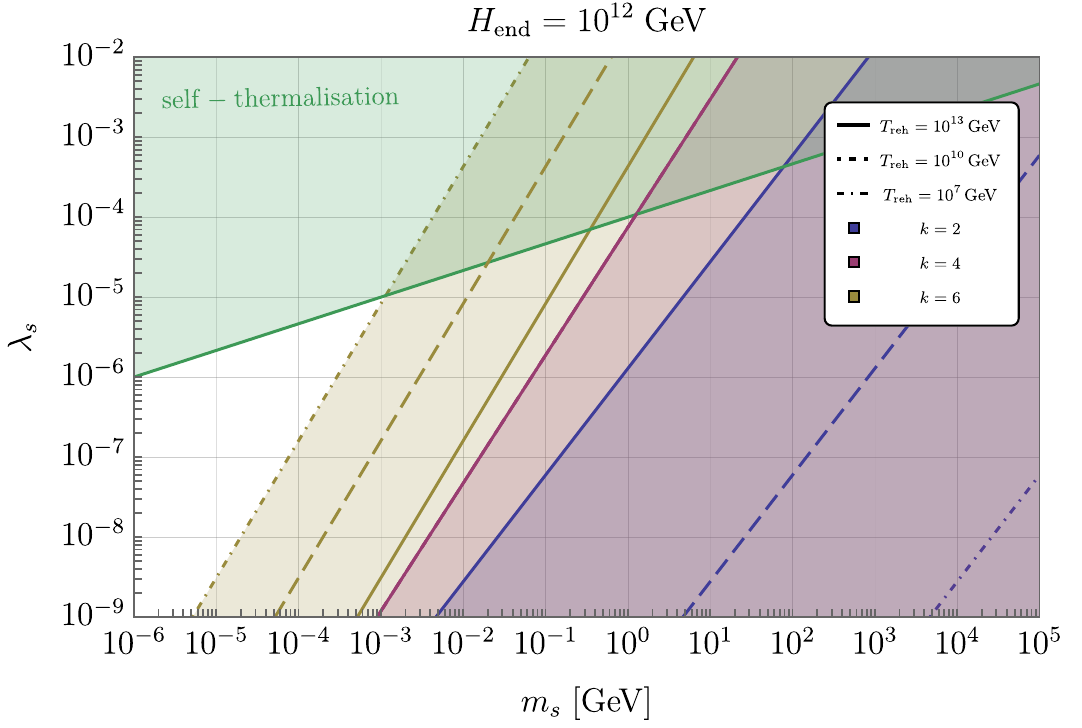}
	\caption{Constraints on the scalar self-interaction coupling $\lambda_s$ in terms of the scalar mass $m_s$ for the $k = 2$ (blue), $k = 4$ (red), and $k = 6$ (olive) cases, with the choice of $g_* = g_{*s} = 100$. With the Hubble parameter at the end of inflation being fixed to be $H_{\rm end} = 10^{12}$ GeV, three values of the reheating temperatures are considered, namely $10^{13}$ GeV (solid), $10^{10}$ GeV (dashed), and $10^7$ GeV (dot-dashed). The self-thermalisation bound (green region) is adopted from Refs.~\cite{Arcadi:2019oxh,Lebedev:2022cic}.}
	\label{fig:inflation_power-law}
\end{figure}
In Fig.~\ref{fig:inflation_power-law}, the constraint on the scalar self-interaction coupling $\lambda_s$ is presented in terms of the scalar mass $m_s$ for the $k = 2$ (blue), $k = 4$ (red), and $k = 6$ (olive) cases, with the choice of $g_* = g_{*s} = 100$. For the Hubble parameter value at the end of inflation, we choose $H_{\rm end} = 10^{12}$ GeV which is the highest value compatible with the latest CMB bound on the tensor-to-scalar ratio, namely $r \leq 0.036$ (95\% C.L.) \cite{Planck:2018jri,BICEP:2021xfz}. In terms of the temperature, this value corresponds to $T \simeq 10^{15}$ GeV. For the Hubble parameter at the end of reheating, three values of the reheating temperatures are considered, $10^{13}$ GeV (solid), $10^{10}$ GeV (dashed), and $10^7$ GeV (dot-dashed). The shaded green region represents the self-thermalisation constraint \cite{Arcadi:2019oxh,Lebedev:2022cic}.
As expected from the discussion above, the bound for the $k = 4$ case is independent of the reheating temperature. In this case, $\Delta_{\rm reh} \to 1$, and we recover the instantaneous reheating scenario as we would have radiation domination all the way from the end of inflation. For $k < 4$, the constraint is proportional to $\Delta_{\rm reh}$, and thus, a finite duration of reheating relaxes the bound. Furthermore, since $\Delta_{\rm reh} \propto H_{\rm reh}^{-1/2} \propto T_{\rm reh}^{-1}$, the bound becomes weaker as the reheating temperature decreases. On the contrary, for $k > 4$, we see an opposite behaviour. The constraint is inversely proportional to $\Delta_{\rm reh}$, meaning that reheating makes the bound stronger. In other words, as the reheating temperature increases, the bound becomes weaker.
One may have an intuitive picture on the results by considering the scaling behaviour of the number density of a decoupled species, $n = n_{\rm end} a_{\rm end}^3 / a^3$. If we convert the scale factor into the cosmic time, the number density scales as $n \propto t^{-3/2}$ during the radiation-dominated era. During the matter-dominated era, the number density behaves as $n \propto t^{-2}$, indicating that the number density in matter domination is diluted with respect to the radiation domination case. On the other hand, during the kination-dominated era, for instance, we have $n \propto t^{-2/9}$, showing the opposite behaviour; the number density decreases slower than in the radiation domination case.

\subsection{Two-stage reheating}
\label{subsec:two-stage-reh}
The reheating phase need not be governed by a single form of the inflaton potential $V_{\rm inf} \propto \phi^k$. During reheating, the inflaton could traverse a generic polynomial potential such as $V_{\rm inf} \simeq \alpha \phi^2 + \beta \phi^4$. As such, in general, the dominant part of the inflaton potential may change its form over the course of reheating. A well-known example is the non-minimally-coupled Higgs inflation model \cite{Futamase:1987ua,Fakir:1990eg,Cervantes-Cota:1995ehs,Komatsu:1999mt,Bezrukov:2007ep}; see, {\it e.g.}, Refs.~\cite{Bezrukov:2008ut,Gong:2015qha,Khan:2025kuh} for the discussion on reheating in this model and Ref.~\cite{Rubio:2018ogq} for a review. 
Motivated by this, in this subsection, we consider a two-stage reheating scenario. The first stage, after the end of inflation, is assumed to be described by the inflaton potential $V_{\rm inf} \propto \phi^{k_1}$, while the second stage is described by $V_{\rm inf} \propto \phi^{k_2}$. The scale factor at the end of the first stage, which we call $a_1$, is thus smaller than the scale factor at the end of reheating $a_{\rm reh}$, which coincides with the scale factor at the end of the second stage $a_2 = a_{\rm reh}$.

One may easily understand that if the reheating terminates before the onset of the scalar $s$ oscillation, we again recover the instantaneous reheating scenario. It is also straightforward to work out the case where we instead have $a_1 < a_{\rm osc} < a_{\rm reh}$. In this case, the ratio of the scale factors $a_{\rm osc}/a_{\rm reh}$ is computed during the second stage, and therefore, the resultant constraint is the same as the one we discussed in Sec.~\ref{subsec:PLreh}; we can simply replace $k$ by $k_2$ in Eq.~\eqref{eqn:PLconstraintEQ}.

The only new case is thus when both the first and second stages end after the onset of the scalar $s$ oscillation, {\it i.e.}, $a_{\rm osc} < a_1 < a_{\rm reh}$. In this case, we have
\begin{align}
	\frac{a_{\rm osc}}{a_{\rm reh}} =
	\frac{a_{\rm osc}}{a_1}
	\frac{a_1}{a_{\rm reh}}
	=
	\left(
	\frac{H_1}{H_{\rm osc}}
	\right)^{\frac{k_1+2}{3k_1}} 
	\left(
	\frac{H_{\rm reh}}{H_1}
	\right)^{\frac{k_2+2}{3k_2}}
	\,,
\end{align}
where $H_1$ is the Hubble parameter at $a = a_1$. Following the same procedure, the yield is obtained as
\begin{align}
	Y \simeq 
	\frac{0.089 g_*^{3/4}}{3^{1/k_1} g_{*s}}
	\alpha_{\rm SY}^{1-\frac{1}{k_1}}
	\lambda_s^{-\frac{1}{k_1}}
	\Delta_1^{1-\frac{4}{k_1}}
	\Delta_2^{1-\frac{4}{k_2}}
	\left(
	\frac{H_{\rm end}}{M_{\rm P}}
	\right)^{\frac{3}{2}}
	\,,
	\label{eqn:TSsyieldgen}
\end{align}
where we have defined
\begin{align}
	\Delta_{1} \equiv
	\sqrt{\frac{H_{\rm end}}{H_1}}
	\,,\qquad
	\Delta_{2} \equiv
	\sqrt{\frac{H_1}{H_{\rm reh}}}
	\,.
\end{align}
Similarly to $\Delta_{\rm reh}$ defined in Sec.~\ref{subsec:PLreh}, $\Delta_1$ ($\Delta_2$) quantifies the duration of the first (second) stage. It is worth noting that when $k_1 = k_2 = k$, we recover the single-stage reheating case discussed in Sec.~\ref{subsec:PLreh}, namely Eq.~\eqref{eqn:PLsyieldgen}.

For the case of the equilibrium probability distribution function of $s$, the yield \eqref{eqn:TSsyieldgen} reduces to
\begin{align}
	Y \simeq 
	C_{k_1} \frac{g_*^{3/4}}{g_{*s}}
	\lambda_s^{-\frac{1}{2k_1}-\frac{1}{2}}
	\Delta_1^{1-\frac{4}{k_1}}
	\Delta_2^{1-\frac{4}{k_2}}
	\left(
	\frac{H_{\rm end}}{M_{\rm P}}
	\right)^{\frac{3}{2}}
	\,,
	\label{eqn:TSsyieldEQ}
\end{align}
where
\begin{align}
	C_{k_1} \equiv 
	2^{\frac{1}{2k_1}-\frac{15}{4}}
	3^{\frac{1}{2}-\frac{3}{2k_1}}
	5^{\frac{1}{4}}
	\pi^{\frac{1}{k_1}-\frac{3}{2}}
	\left[
	\frac{\Gamma(3/4)}{\Gamma(1/4)}
	\right]^{1-\frac{1}{k_1}}
	\,.
	\label{eqn:Ck1-definition}
\end{align}
We stress that the coefficient $C_{k_1}$ and the exponent of $\lambda_s$ depend only on $k_1$.
We then find the following constraint from Eq.~\eqref{eqn:abundance-constraint}:
\begin{align}
	m_s \lambda_s^{-\frac{1}{2k_1}-\frac{1}{2}}
	\lesssim 
	\frac{4.4\times 10^{-10}}{C_{k_1}}
	\left(
	\frac{g_{*s}}{g_*^{3/4}}
	\right)
	\Delta_1^{\frac{4}{k_1}-1}
	\Delta_2^{\frac{4}{k_2}-1}
	\left(
	\frac{M_{\rm P}}{H_{\rm end}}
	\right)^{\frac{3}{2}}
	\, {\rm GeV}
	\,.
	\label{eqn:TSconstraintEQ}
\end{align}
The result for the non-equilibrium case is reported in Appendix~\ref{apdx:non-Eq}.

\begin{figure}
    \centering
    \includegraphics[width=1\linewidth]{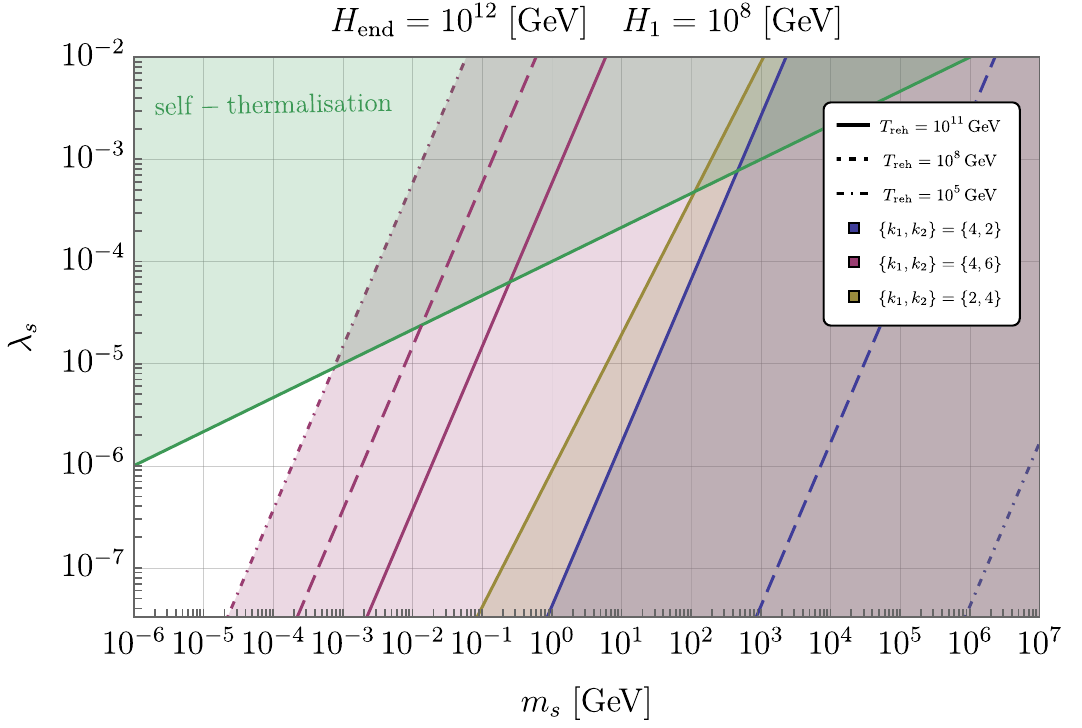}
    \caption{Constraints on the scalar self-interaction coupling $\lambda_s$ in terms of the scalar mass $m_s$ for the two-stage reheating scenario. Three cases of $\{k_1, k_2\}$ are considered: $\{4, 2\}$ (blue), $\{4, 6\}$ (red), and $\{2, 4\}$ (olive). As in Fig.~\ref{fig:inflation_power-law}, the Hubble parameter at the end of inflation is chosen to be $H_{\rm end} = 10^{12}$ GeV, and $g_* = g_{*s} = 100$ is taken. The reheating temperature $T_{\rm reh}$ is varied, while fixing the Hubble parameter at the end of the first stage to $H_1 = 10^8$ GeV. The green region represents the self-thermalisation bound \cite{Arcadi:2019oxh,Lebedev:2022cic}.}
    \label{fig:2stage_H1_fixed}
\end{figure}
Figure~\ref{fig:2stage_H1_fixed} shows the constraint on the scalar self-interaction coupling $\lambda_s$ in terms of the scalar mass $m_s$ for the two-stage reheating scenario, Eq.~\eqref{eqn:TSconstraintEQ}. The Hubble parameter at the end of inflation (the end of the first stage) is set to $H_{\rm end} = 10^{12}$ GeV ($H_1 = 10^8$ GeV), and $g_* = g_{*s} = 100$ is taken. For the inflaton potential shapes during the first and second stages, three cases are considered, $\{k_1, k_2\} = \{4, 2\}$ (blue), $\{4, 6\}$ (red), and $\{2, 4\}$ (olive). In addition, three different values are considered for the reheating temperature, $T_{\rm reh} = 10^{11}$ GeV (solid), $10^{8}$ GeV (dashed), and $10^{5}$ GeV (dot-dashed). One may explicitly see that the $\{k_1, k_2\} = \{2, 4\}$ case is independent of the reheating temperature. This is because of the fixed $H_1$ and the choice of $k = 4$ for the second stage, which, as we saw earlier, behaves like the instantaneous reheating case.
It is worth noting that the choice of the $y$-axis in Fig.~\ref{fig:2stage_H1_fixed}, namely $\lambda_s \simeq 10^{-8}$, comes from the triviality condition, $a_1 > a_{\rm osc}$ or, equivalently, $H_1 < H_{\rm osc}$.\footnote{
The $a_{1}< a_{\rm osc}$ case reduces to the single-stage reheating scenario as we discussed earlier.
} Since $H_{\rm osc} \simeq \sqrt{3 \alpha_{\rm SY} \lambda_s} H_{\rm end}$, we get, for $H_{\rm{end}} \simeq 10^{12}$ GeV and $H_1 \simeq 10^8$ GeV, $\lambda_s \gtrsim 10^{-8}$ in the equilibrium case. In addition, $H_1 > H_{\rm reh}$ is required by definition in the scenario under consideration.
From Fig.~\ref{fig:2stage_H1_fixed}, we see that in the $\{k_1, k_2\} = \{4, 2\}$ case, the longer it takes for reheating to finish, the weaker the bound becomes. This is essentially due to the increase of the matter-dominated epoch, which dilutes the gravitationally-produced relics. On the other hand, for the $\{k_1, k_2\} = \{4, 6\}$ case, the bound becomes stronger as the end of reheating gets delayed. As we discussed earlier, this is because prolonging the kination phase enhances the number density of relics compared to the radiation-dominated case.

\begin{figure}
    \centering
    \includegraphics[width=1\linewidth]{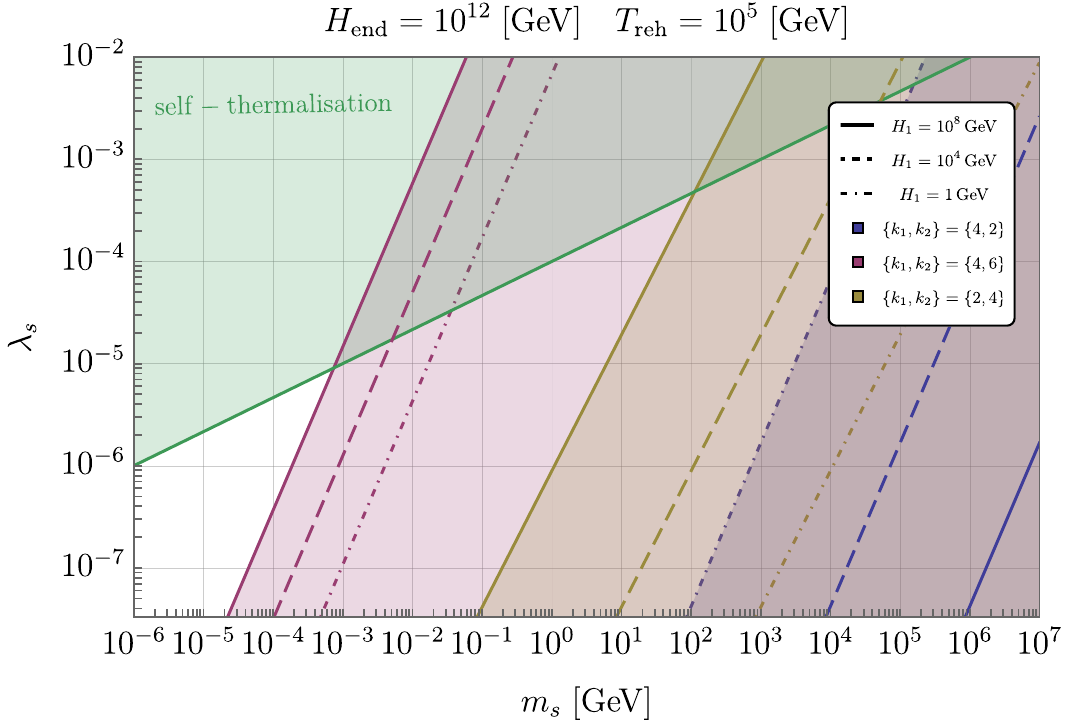}
    \caption{Constraints on the scalar self-interaction coupling $\lambda_s$ in terms of the scalar mass $m_s$ for the two-stage reheating scenario. As in Fig.~\ref{fig:2stage_H1_fixed}, three cases of $\{k_1, k_2\}$, namely $\{4, 2\}$ (blue), $\{4, 6\}$ (red), and $\{2, 4\}$ (olive), are considered, the Hubble parameter at the end of inflation is chosen to be $H_{\rm end} = 10^{12}$ GeV, and $g_* = g_{*s} = 100$ is taken. The Hubble parameter at the end of the first stage, $H_1$, is varied, while fixing the reheating temperature to $T_{\rm reh} = 10^5$ GeV. The green region represents the self-thermalisation bound \cite{Arcadi:2019oxh,Lebedev:2022cic}.}
    \label{fig:2stage_H1_changes}
\end{figure}
Instead of fixing $H_1$ and varying $T_{\rm reh}$, in Fig.~\ref{fig:2stage_H1_changes}, $H_1$ is varied with the reheating temperature fixed to $T_{\rm reh} = 10^5$ GeV. We see in this case that the bound for the $\{k_1, k_2\} = \{2, 4\}$ case changes accordingly. Decreasing $H_1$ prolongs the length of the matter-dominated era, and thus, the bound gets weakened. The $\{k_1, k_2\} = \{4, 6\}$ case follows the same tendency. This is because the kination epoch becomes shortened as $H_1$ decreases. On the other hand, the $\{k_1, k_2\} = \{4, 2\}$ case shows the opposite behaviour; the bound becomes stronger as we decrease $H_1$ as the matter-dominated epoch lasts shorter.

\subsection{$m$-stage reheating}
\label{subsec:m-stage-reh}
We can generalise the two-stage reheating scenario to an $m$-stage reheating scenario in a straightforward manner. We again focus on the case where all the $m$ stages end after the onset of the scalar $s$ oscillation, {\it i.e.}, $a_{\rm osc} < a_1 < a_2 < \cdots < a_m = a_{\rm reh}$. Repeating the analysis presented in Sec.~\ref{subsec:two-stage-reh}, we arrive at the following expression for the yield:
\begin{align}
	Y \simeq 
	\frac{0.089 g_*^{3/4}}{3^{1/k_1} g_{*s}}
	\alpha_{\rm SY}^{1-\frac{1}{k_1}}
	\lambda_s^{-\frac{1}{k_1}}
	\Delta_1^{1-\frac{4}{k_1}}
	\Delta_2^{1-\frac{4}{k_2}}
	\cdots
	\Delta_m^{1-\frac{4}{k_m}}
	\left(
	\frac{H_{\rm end}}{M_{\rm P}}
	\right)^{\frac{3}{2}}
	\,,
\end{align}
where
\begin{align}
	\Delta_{1} \equiv
	\sqrt{\frac{H_{\rm end}}{H_1}}
	\,,\quad
	\Delta_{2} \equiv
	\sqrt{\frac{H_1}{H_2}}
	\,,\quad\cdots\,,\quad
	\Delta_{m} \equiv
	\sqrt{\frac{H_{m-1}}{H_{\rm reh}}}
	\,.
\end{align}
The abundance constraint \eqref{eqn:abundance-constraint}, for the case of the equilibrium probability distribution function of $s$, then becomes
\begin{align}
	m_s \lambda_s^{-\frac{1}{2k_1}-\frac{1}{2}}
	\lesssim 
	\frac{4.4\times 10^{-10}}{C_{k_1}}
	\left(
	\frac{g_{*s}}{g_*^{3/4}}
	\right)
	\Delta_1^{\frac{4}{k_1}-1}
	\Delta_2^{\frac{4}{k_2}-1}
	\cdots 
	\Delta_m^{\frac{4}{k_m}-1}
	\left(
	\frac{M_{\rm P}}{H_{\rm end}}
	\right)^{\frac{3}{2}}
	\, {\rm GeV}
	\,,
\end{align}
where the coefficient $C_{k_1}$ is given in Eq.~\eqref{eqn:Ck1-definition}. We stress that the coefficient $C_{k_1}$ as well as the exponent of $\lambda_s$ depend again only on $k_1$.

\section{Production during reheating}
\label{sec:prodreh}
We now consider the production of the scalar $s$ during reheating. Since the scalar $s$ is assumed to be completely decoupled, {\it i.e.}, no interaction between the scalar $s$ and other fields is assumed, the only possible production is through gravity.
At high energy scales such as our early Universe, we expect quantum gravity to play a non-negligible role. Considering gravitational interactions between the inflaton and the scalar $s$, one may parametrise the effect of quantum gravity in an effective field theory framework through the following Planck-suppressed operators \cite{Lebedev:2022cic,Lebedev:2022ljz}:
\begin{align}
	\mathcal{L}_{\rm QG} \supset
	\frac{C_1}{M_{\rm P}^2}(\partial s)^2 \phi^2
	+\frac{C_2}{M_{\rm P}^2} \phi^4 s^2
	+\frac{C_3}{M_{\rm P}^2} \phi^2 s^4
	\,,
\end{align}
where $C_1$, $C_2$, and $C_3$ are the Wilson coefficients. Here, we have imposed $Z_2$ symmetry and assumed that the inflaton field value is larger than the inflaton mass. The first two operators, namely
\begin{align}
	\mathcal{O}_1 =
	\frac{C_1}{M_{\rm P}^2} (\partial s)^2 \phi^2
	\,,\qquad
	\mathcal{O}_2 =
	\frac{C_2}{M_{\rm P}^2} \phi^4 s^2
	\,,
\end{align}
are responsible for the pair production of the scalar $s$, and effects of these quantum gravity-induced operators on the production of a scalar have been thoroughly investigated in Ref.~\cite{Lebedev:2022cic} for the quadratic and quartic inflaton potential cases.
It is worth mentioning that the third operator, $C_3 \phi^2 s^4 / M_{\rm P}^2$, is less consequential and subdominant compared to the operator $\mathcal{O}_2$; see Ref.~\cite{Lebedev:2022ljz} for more details.
In this work, we generalise the analysis of Ref.~\cite{Lebedev:2022cic} and study the production of $s$ during the reheating phase with a generic power-law inflaton potential as well as in the case of a multi-stage reheating scenario.
We note that the operator $\mathcal{O}_1$, under the assumption that General Relativity still holds at the relevant reheating energy scale, could be generated via the graviton exchange \cite{Mambrini:2021zpp,Lebedev:2022ljz,Lebedev:2022cic}; in such a case, the Wilson coefficient is fixed to be $C_1 \simeq 0.25/(k+2)$ for the inflaton potential $V_{\rm inf}(\phi) \propto \phi^k$.

Particle production through the inflaton oscillation during the post-inflationary phase has been extensively studied in the literature; see, {\it e.g.}, Refs.~\cite{Abbott:1982hn,Ichikawa:2008ne,Nurmi:2015ema,Garcia:2020wiy}. We closely follow the notations of Ref.~\cite{Garcia:2020wiy}. For concreteness, let us take the inflaton to oscillate around the potential minimum of the form
\begin{align}
	V_{\rm inf}(\phi) = \lambda\frac{\phi^k}{M_{\rm P}^{k-4}}
	\,,
	\label{eqn:inf-potential}
\end{align}
after the end of inflation. Throughout this section, we take $k \geq 2$.
During the reheating phase, the inflaton can be described as
\begin{align}
	\phi(t) =
	\phi_0(t) \sum_{\ell=-\infty}^\infty \mathcal{P}_\ell e^{-i\ell\omega t}
	\,,
\end{align}
where $\phi_0(t)$ is the envelope, the part $\sum_\ell \mathcal{P}_\ell e^{-i\ell\omega t}$ represents the short time-scale oscillations, and the oscillation frequency $\omega$ is given by
\begin{align}
	\omega =
	m_\phi
	\sqrt{\frac{\pi k}{2(k-1)}}
	\frac{\Gamma(1/2+1/k)}{\Gamma(1/k)}
	\,,
\end{align}
with $m_\phi$ being the inflaton mass. For the inflaton potential \eqref{eqn:inf-potential}, from $m_\phi^2 = V''(\phi_0)$, we read
\begin{align}
	\omega = 
	\frac{\sqrt{\pi \lambda} k}{\sqrt{2}}
	\frac{\phi_0^{(k-2)/2}}{M_{\rm P}^{(k-4)/2}}
	\left[
	\frac{\Gamma(1/2+1/k)}{\Gamma(1/k)}
	\right]
	\,.
	\label{eqn:inf-osc-freq}
\end{align}

\subsection{Operator 1}
Let us first investigate the production of the scalar $s$ through the operator
\begin{align}
    \mathcal{O}_1 =
	\frac{C_1}{M_{\rm P}^2}
	\left(\partial s\right)^2 \phi^2
	\,.
\end{align}
The amplitude for the interaction between the $\ell$-th mode of the inflaton condensate and the scalar $s$ is given by
\begin{align}
	\mathcal{M}_\ell =
	i \frac{2 C_1}{M_{\rm P}^2}
	\left( p \cdot q \right)
	\phi_0^2 \left(\mathcal{P}^2\right)_\ell
	\,,
\end{align}
where the product of the momenta, $p \cdot q$, is due to the derivatives of the scalar $s$ in the operator. In the limit where the final-state $s$ scalar mass is negligible, the momentum factor can be replaced by the energy of the each oscillation mode of the inflaton condensate, $E_\ell = \ell \omega$, as
\begin{equation}
    p \cdot q =
    \frac{E_\ell^2}{2} = 
    \frac{\ell^2 \omega^2}{2}
    \,.
\end{equation}
Therefore, we read
\begin{align}
    |\mathcal{M}_\ell|^2 =
    \frac{C_1^2 \ell^4 \omega^4}{M_{\rm P}^4} 
    \phi_0^4 |\left(\mathcal{P}^2\right)_\ell|^2
    \,.
\end{align}
The number density production rate is then given by
\begin{align}
	\Gamma_{1}
	= \frac{C_1^2 \phi_0^4 \omega^4}{16 \pi (1 + w_\phi) \rho_\phi M_{\rm P}^4}
	\sum_{\ell = 1}^{\infty}
	\ell^4 |\left(\mathcal{P}^2\right)_\ell|^2
	\,,
\end{align}
where we have taken into account the symmetry factor of $1/2$ for the final state, and $w_\phi = (k-2)/(k+2)$ is the equation of state, averaged over oscillations. The Boltzmann equation for the number density $n_s$ of the scalar $s$ is then given by
\begin{align}
	\dot{n}_s + 3Hn_s = 
	2 \Gamma_1 (1 + w_\phi) \rho_\phi
	\,,
\end{align}
where the factor $2$ comes from the pair production of $s$ per interaction. We thus find
\begin{align}
	\dot{n}_s + 3Hn_s = 
	\frac{C_1^2 \phi_0^4 \omega^4}{8 \pi M_{\rm P}^4}
	\sum_\ell \ell^4 |(\mathcal{P}^2)_\ell|^2
	\,,
\end{align}
or, equivalently,
\begin{align}
	\frac{d}{dt}\left( n_s a^3 \right) =
	\frac{a^3 C_1^2 \phi_0^4 \omega^4}{8 \pi M_{\rm P}^4}
	\sum_\ell \ell^4 |(\mathcal{P}^2)_\ell|^2
	\,.
	\label{eqn:O1BE-singlestage}
\end{align}
Utilising Eq.~\eqref{eqn:inf-osc-freq} for the oscillation frequency $\omega$, the Boltzmann equation becomes
\begin{align}
	\frac{d}{dt}\left( n_s a^3 \right) =
	\frac{\pi k^4 \lambda^2 a^3 C_1^2 \phi_0^{2k}}{32 M_{\rm P}^{2(k-2)}}
	\left[
	\frac{\Gamma(1/2+1/k)}{\Gamma(1/k)}
	\right]^4
	\sum_\ell \ell^4 |(\mathcal{P}^2)_\ell|^2
	\,.
\end{align}

In order to integrate the Boltzmann equation, we first change the integration variable from the cosmic time to the scale factor using $dt = (aH)^{-1} da$ and then note that the amplitude of $\phi_0$ evolves as $\phi_0 \propto  a^{-6/(k+2)}$. Denoting $\phi_0(a = a_0) = \phi_{\rm end}$ and setting $a_0 = 1$, we have
\begin{align}
	n_s a^3 - n_s(a_0) &=
	\frac{\pi k^4 \lambda^2 C_1^2 \phi_{\rm end}^{2k}}{32 M_{\rm P}^{2(k-2)} H_{\rm end}}
	\left[
	\frac{\Gamma(1/2+1/k)}{\Gamma(1/k)}
	\right]^4
	P
	\int_1^a da' \, (a')^{\frac{4-7k}{k+2}} 
	\nonumber\\&=
	\frac{\pi k^4 (k+2) \lambda^2 C_1^2 \phi_{\rm end}^{2k}}{192 (k-1) M_{\rm P}^{2(k-2)} H_{\rm end}}
	\left[
	\frac{\Gamma(1/2+1/k)}{\Gamma(1/k)}
	\right]^4
	P
	\left[
	1-a^{\frac{6(1-k)}{k+2}}
	\right]
	\,,
\end{align}
where $P \equiv \sum_\ell \ell^4 |(\mathcal{P}^2)_\ell|^2$. The computation of $P$ and $(\mathcal{P}^\gamma)_\ell$ is presented in Appendix \ref{apdx:compP}.
One may notice that the exponent $6(1-k)/(k+2)$ is always negative as we are interested in $k \geq 2$ cases. It indicates that at a sufficiently late time, the scale factor term drops out. In other words, the major contribution to the number density of the scalar $s$ comes from the earliest time of reheating. The produced abundance of $s$ then evolves ``freely'' like any decoupled species. It also makes the comparison between the production during the inflationary phase and the production during the reheating phase simple, because both of them can be seen to be fixed at the end of inflation and then evolve in the same way as $a^{-3}$. Therefore, unless fine-tuned, one of the two production mechanisms will dominate. In Sec.~\ref{sec:prodinf}, we focused on the case where the production during inflation dominates. In this section, we study the case where the production during reheating dominates.

Taking the late-time limit of $a \gg 1$, we get
\begin{align}
	n_s \simeq 
	\frac{\pi k^4 (k+2) \lambda^2 C_1^2 \phi_{\rm end}^{2k}}{192 (k-1) M_{\rm P}^{2(k-2)} H_{\rm end} a^3}
	\left[
	\frac{\Gamma(1/2+1/k)}{\Gamma(1/k)}
	\right]^4
	P
	\,.
\end{align}
The inflaton field value at the end of inflation can be approximated as
\begin{align}
	\phi_{\rm end} \simeq
	\left(
	\frac{3}{\lambda}
	\right)^{\frac{1}{k}} \left(
	\frac{H_{\rm end}}{M_{\rm P}}
	\right)^{\frac{2}{k}} M_{\rm P}
	\,,
\end{align}
and thus, we find
\begin{align}
	n_s \simeq
	\frac{3 \pi k^4 (k+2) C_1^2}{64 (k-1) a^3}
	H_{\rm end}^3
	\left[
	\frac{\Gamma(1/2+1/k)}{\Gamma(1/k)}
	\right]^4
	P
	\,.
\end{align}

The yield at the end of reheating is then given by
\begin{align}
	Y = \frac{n_s}{s_{\rm SM}} &\simeq
	\frac{135 k^4 (k+2) C_1^2}{128 \pi (k-1) g_{*s}}
	\frac{H_{\rm end}^3}{T_{\rm reh}^3 a_{\rm reh}^3}
	\left[
	\frac{\Gamma(1/2+1/k)}{\Gamma(1/k)}
	\right]^4
	P
	\,,
\end{align}
where we have used $s_{\rm SM} = (2\pi^2/45)g_{*s}T^3$.
From $H = H_{\rm end} \; a^{-3k/(k+2)}$, we may write
\begin{align}
	a_{\rm reh} = \left(
	\frac{H_{\rm end}}{H_{\rm reh}}
	\right)^{\frac{k+2}{3k}}
	\,.
\end{align}
Moreover, using $3 M_{\rm P}^2 H_{\rm reh}^2 = (\pi^2 g_{*}/30) T_{\rm reh}^4$, the yield can be expressed as
\begin{align}
	Y &\simeq
	0.064 C_1^2
	\frac{k^4 (k+2)}{k-1}
	\left(\frac{g_*^{3/4}}{g_{*s}}\right)
	\left[
	\frac{\Gamma(1/2+1/k)}{\Gamma(1/k)}
	\right]^4
	P \;
	\Delta_{\rm reh}^{1-\frac{4}{k}}
	\left(
	\frac{H_{\rm end}}{M_{\rm P}}
	\right)^{\frac{3}{2}}
	\,.
	\label{eqn:yield-O1-single}
\end{align}
The abundance constraint \eqref{eqn:abundance-constraint} then translates to the following constraint on the Wilson coefficient $C_1$:
\begin{align}
	|C_1|
	\lesssim 8.3 \times 10^{-5}
	\sqrt{\frac{k-1}{k^4 (k+2)}}
	\left[
	\frac{\Gamma(1/k)}{\Gamma(1/2+1/k)}
	\right]^2
	\sqrt{\frac{g_{*s}}{g_*^{3/4}}}
	P^{-\frac{1}{2}}
	\Delta_{\rm reh}^{\frac{4-k}{2k}}
	\left(
	\frac{M_{\rm P}}{H_{\rm end}}
	\right)^{\frac{3}{4}}
	\,\sqrt{\frac{{\rm GeV}}{m_s}}
	\,.
	\label{eqn:C1constraint}
\end{align}
Explicitly, expressions of the constraint \eqref{eqn:C1constraint} for the $k = 2$, $k = 4$, and $k = 6$ cases are given by
\begin{align}
	|C_1|
	&\lesssim 
	3.3 \times 10^{-5}
	\sqrt{\frac{g_{*s}}{g_*^{3/4}}}
	\Delta_{\rm reh}^{\frac{1}{2}}
	\left(
	\frac{M_{\rm P}}{H_{\rm end}}
	\right)^{\frac{3}{4}}
	\,\sqrt{\frac{{\rm GeV}}{m_s}}
	&\quad \text{for } k = 2
	\,,\label{eqn:C1constraintk2}\\
	|C_1|
	&\lesssim 
	3.1 \times 10^{-5}
	\sqrt{\frac{g_{*s}}{g_*^{3/4}}}
	\left(
	\frac{M_{\rm P}}{H_{\rm end}}
	\right)^{\frac{3}{4}}
	\,\sqrt{\frac{{\rm GeV}}{m_s}}
	&\quad \text{for } k = 4
	\,,\label{eqn:C1constraintk4}\\
	|C_1|
	&\lesssim 
	2.7 \times 10^{-5}
	\sqrt{\frac{g_{*s}}{g_*^{3/4}}}
	\Delta_{\rm reh}^{-\frac{1}{6}}
	\left(
	\frac{M_{\rm P}}{H_{\rm end}}
	\right)^{\frac{3}{4}}
	\,\sqrt{\frac{{\rm GeV}}{m_s}}
	&\quad \text{for } k = 6
	\,,
\end{align}
where we have used $P = 1$, $P \approx 1.11$, and $P \approx 1.29$ for $k = 2$, $k = 4$, and $k = 6$, respectively. We note that Eqs.~\eqref{eqn:C1constraintk2} and \eqref{eqn:C1constraintk4} agree with the constraints reported in Ref.~\cite{Lebedev:2022cic}.

\begin{figure}
	\centering
	\includegraphics[width=1\linewidth]{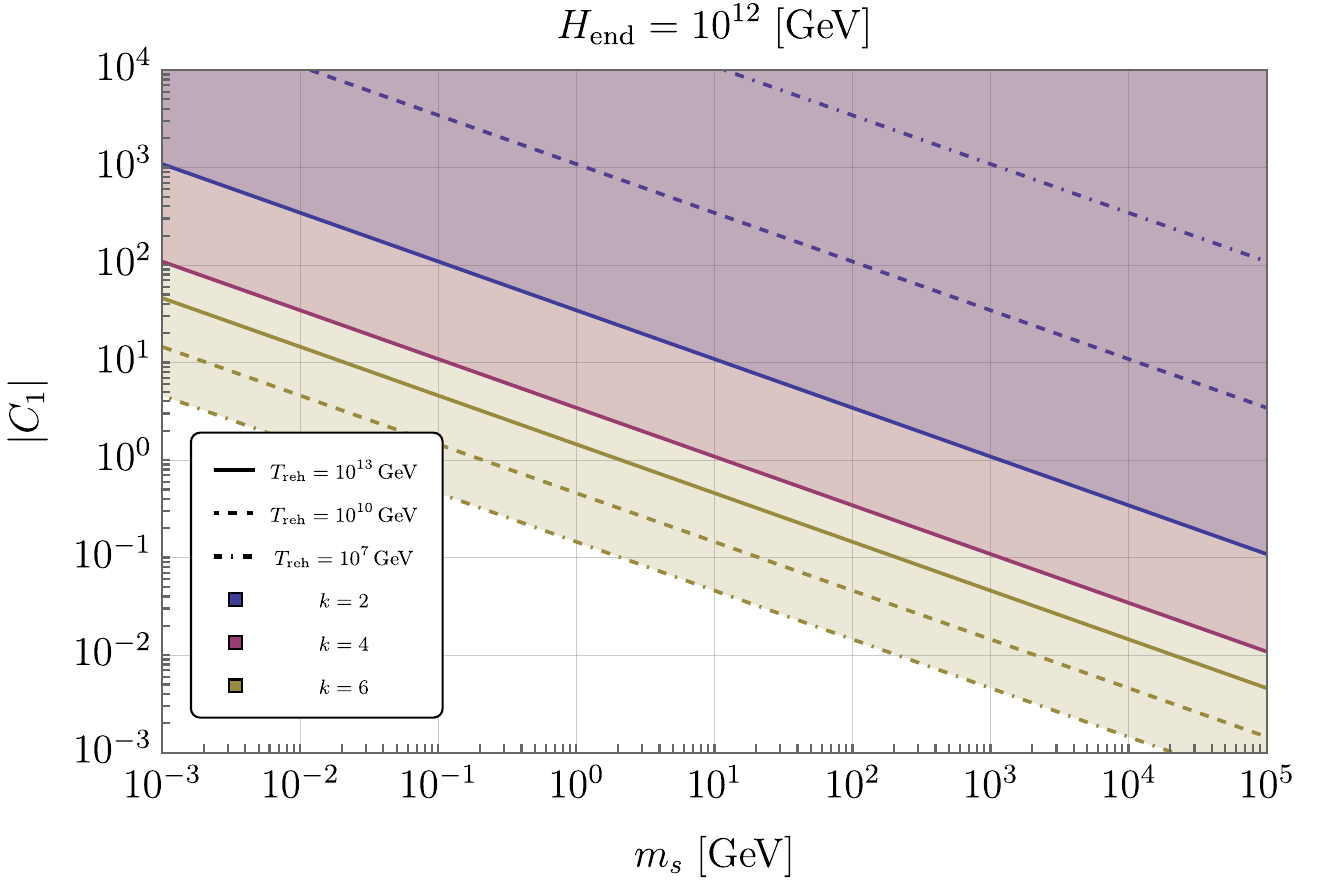}
	\caption{Upper limits on $|C_1|$ in terms of the scalar mass $m_s$ for the $k = 2$ (blue), $k = 4$ (red), and $k = 6$ (olive) cases. Three different reheating temperatures are considered, $10^{13}$ GeV (solid), $10^{10}$ GeV (dashed), and $10^7$ GeV (dot-dashed). The Hubble parameter at the end of inflation is fixed to be $H_{\rm end} = 10^{12}$ GeV, and $g_* = g_{*s} = 100$ is chosen.}
	\label{fig:C1_reheating}
\end{figure}
We present the constraints on $C_1$ for $k = 2$ (blue), $k = 4$ (red), and $k = 6$ (olive) cases in Fig.~\ref{fig:C1_reheating} for three reheating temperature values, $10^{13}$ GeV (solid), $10^{10}$ GeV (dashed), and $10^7$ GeV (dot-dashed). The Hubble parameter at the end of inflation is fixed to be $H_{\rm end} = 10^{12}$ GeV, and $g_* = g_{*s} = 100$ is chosen. The different lines indicate the upper limit of $|C_1|$. One may notice that for the $k = 2$ case, the bound quickly weakens as the reheating temperature gets lowered, while for the $k = 6$ case, the bound becomes stronger instead. For the $k = 4$ case, the constraint is independent of the reheating temperature. This behaviour is analogous to the one in the gravitational production during inflation discussed in Sec.~\ref{sec:prodinf}. We see from Fig.~\ref{fig:C1_reheating} that it is possible for $|C_1|$ to take arguably the most natural value of $\mathcal{O}(0.01-1)$ in the GeV--TeV mass range. We also observe that the parameter space opens up to a much wider mass range with a lower reheating temperature.

\subsection{Operator 2}
The production of the scalar $s$ through the operator
\begin{align}
    \mathcal{O}_2 =
	\frac{C_2}{M_{\rm P}^2} \phi^4 s^2
	\,,
\end{align}
can be computed in the same manner. The amplitude-squared for the interaction between the $\ell$-th mode of the inflaton condensate and the scalar $s$ described by $\mathcal{O}_2$ is given by
\begin{align}
    |\mathcal{M}_\ell|^2 =
    \frac{4 C_2^2}{M_{\rm P}^4} 
    \phi_0^8 |\left(\mathcal{P}^4\right)_\ell|^2
    \,.
\end{align}
Then, the number density production rate is given by
\begin{align}
	\Gamma_{2} = 
	\frac{C_2^2 \phi_0^8}{4\pi (1 + w_\phi) \rho_\phi M_{\rm P}^4}
	\sum_{\ell=1}^{\infty} 
	|\left(\mathcal{P}^4\right)_\ell|^2
	\,,
\end{align}
where we have taken into account the symmetry factor of $1/2$ for the final state, and the negligible $s$ mass has been dropped. The Boltzmann equation for the number density $n_s$ of the scalar $s$ reads
\begin{align}
	\dot{n}_s + 3Hn_s = 
	2\Gamma_2(1 + w_\phi)\rho_\phi
	\,,
\end{align}
or, equivalently,
\begin{align}
	\frac{d}{dt}\left( n_s a^3 \right) =
	\frac{a^3 C_2^2 \phi_0^8}{2 \pi M_{\rm P}^4}
	\sum_\ell \ell |(\mathcal{P}^4)_\ell|^2
	\,.
	\label{eqn:O2BE-singlestage}
\end{align}
Following the same procedure as before, we integrate the Boltzmann equation to obtain
\begin{align}
	n_s a^3 &=
	\frac{(k+2) C_2^2 \phi_{\rm end}^8}{12\pi (7-k) M_{\rm P}^4 H_{\rm end}} Q
	\left[
	1-a^{-\frac{42-6k}{k+2}}
	\right]
	\,,
\end{align}
where $Q \equiv \sum_\ell \ell |(\mathcal{P}^4)_\ell|^2$. The computation of $Q$ and $(\mathcal{P}^\gamma)_\ell$ is presented in Appendix \ref{apdx:compP}. We see that the exponent of the scale factor $a$, namely $(6k-42)/(k+2)$, is always negative as long as $k$ is less than 7. As a consequence, as in the case of $\mathcal{O}_1$, the scale factor term can be discarded in the late-time limit. When $k$ is larger than 7, however, the number density of the produced scalar instead grows with time, and thus, it becomes IR-dominated rather than UV-dominated. We do not consider $k \geq 7$ cases further in this work. Taking the late-time limit of $a \gg 1$, the number density takes
\begin{align}
	n_s \simeq 
	\frac{(k+2) C_2^2 \phi_{\rm end}^8}{12\pi (7-k) M_{\rm P}^4 H_{\rm end}a^3} Q
	\,.
\end{align}

The yield, $Y = n_s / s_{\rm SM}$, at the end of reheating is then given by
\begin{align}
	Y \simeq
	0.012 C_2^2
	\left(\frac{k+2}{7-k}\right)
	\left(
	\frac{g_*^{3/4}}{g_{*s}}
	\right)
	Q
	\Delta_{\rm reh}^{1 - \frac{4}{k}}
	\left(
	\frac{M_{\rm P}}{H_{\rm end}}
	\right)^{\frac{5}{2}}
	\left(
	\frac{\phi_{\rm end}}{M_{\rm P}}
	\right)^8
	\,,
	\label{eqn:yield-O2-single}
\end{align}
and the abundance constraint \eqref{eqn:abundance-constraint} translates to an upper bound of the Wilson coefficient $C_2$ as follows:
\begin{align}
	|C_2|
	\lesssim 
	1.9 \times 10^{-4}
	\sqrt{\frac{7-k}{k+2}}
	\sqrt{\frac{g_{*s}}{g_*^{3/4}}}
	Q^{-\frac{1}{2}}
	\Delta_{\rm reh}^{\frac{4-k}{2k}}
	\left(
	\frac{H_{\rm end}}{M_{\rm P}}
	\right)^{\frac{5}{4}}
	\left(
	\frac{M_{\rm P}}{\phi_{\rm end}}
	\right)^4
	\sqrt{\frac{{\rm GeV}}{m_s}}
	\,.
	\label{eqn:C2constraint}
\end{align}
Explicitly, expressions of the constraint \eqref{eqn:C2constraint} for the $k = 2$, $k = 4$, and $k = 6$ cases are given by
\begin{align}
	|C_2|
	&\lesssim 
	5.8 \times 10^{-4}
	\sqrt{\frac{g_{*s}}{g_*^{3/4}}}
	\Delta_{\rm reh}^{\frac{1}{2}}
	\left(
	\frac{H_{\rm end}}{M_{\rm P}}
	\right)^{\frac{5}{4}}
	\left(
	\frac{M_{\rm P}}{\phi_{\rm end}}
	\right)^4
	\sqrt{\frac{{\rm GeV}}{m_s}}
	&\quad \text{for } k = 2
	\,,\label{eqn:C2constraintk2}\\
	|C_2|
	&\lesssim 
	3.7 \times 10^{-4}
	\sqrt{\frac{g_{*s}}{g_*^{3/4}}}
	\left(
	\frac{H_{\rm end}}{M_{\rm P}}
	\right)^{\frac{5}{4}}
	\left(
	\frac{M_{\rm P}}{\phi_{\rm end}}
	\right)^4
	\sqrt{\frac{{\rm GeV}}{m_s}}
	&\quad \text{for } k = 4  \,,\label{eqn:C2constraintk4}\\
	|C_2|
	&\lesssim 
	1.9 \times 10^{-4}
	\sqrt{\frac{g_{*s}}{g_*^{3/4}}}
	\Delta_{\rm reh}^{-\frac{1}{6}}
	\left(
	\frac{H_{\rm end}}{M_{\rm P}}
	\right)^{\frac{5}{4}}
	\left(
	\frac{M_{\rm P}}{\phi_{\rm end}}
	\right)^4
	\sqrt{\frac{{\rm GeV}}{m_s}}
	&\quad \text{for } k = 6
	\,,
\end{align}
where we have used $Q = 9/64$, $Q \approx 0.141$, and $Q \approx 0.140$ for $k = 2$, $k = 4$, and $k = 6$, respectively. We note that Eqs.~\eqref{eqn:C2constraintk2} and \eqref{eqn:C2constraintk4} recover the constraints given in Ref.~\cite{Lebedev:2022cic}.

\begin{figure}
	\centering
	\includegraphics[width=1\linewidth]{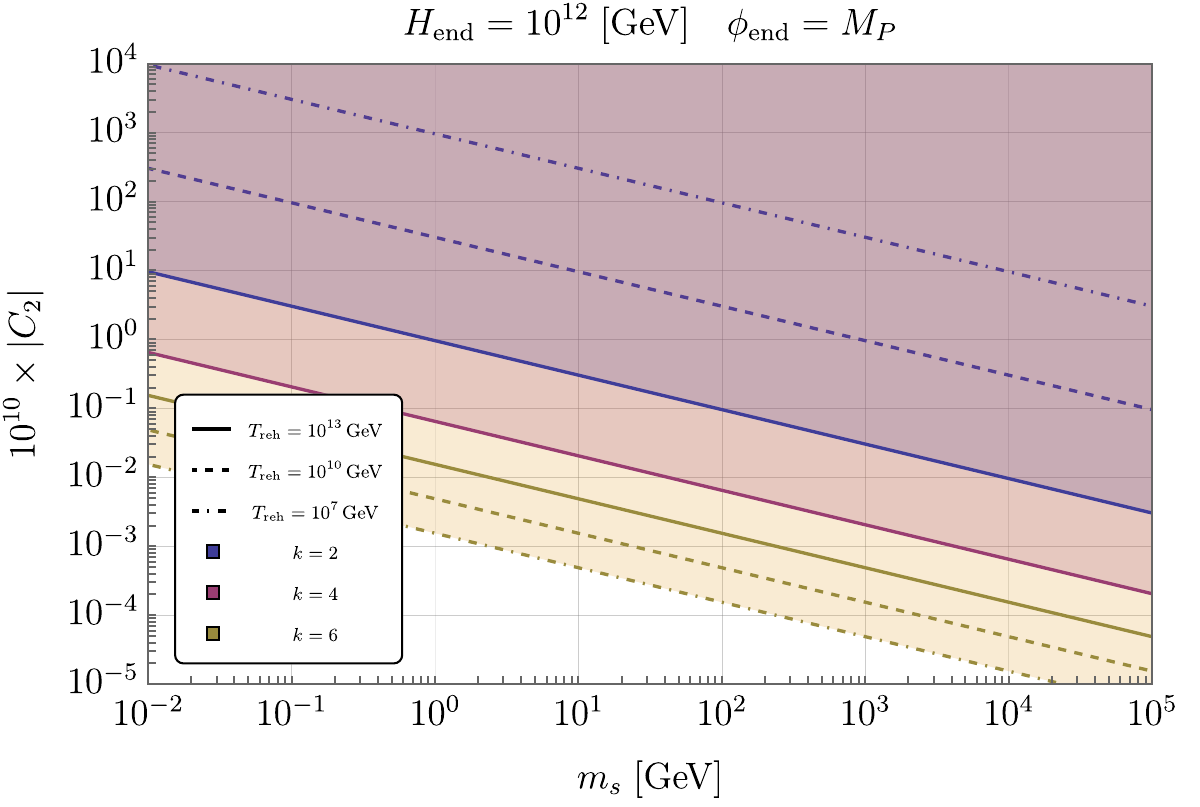}
	\caption{Upper limits on $ |C_2|$ in terms of the scalar mass $m_s$ for the $k = 2$ (blue), $k = 4$ (red), and $k = 6$ (olive) cases. As in Fig.~\ref{fig:C1_reheating}, three different reheating temperatures are considered, $10^{13}$ GeV (solid), $10^{10}$ GeV (dashed), and $10^7$ GeV (dot-dashed), with $g_* = g_{*s} = 100$, $\phi_{\rm end} = M_{\rm P}$, and $H_{\rm end} = 10^{12}$ GeV.}
	\label{fig:C2_reheating}
\end{figure}
The constraints on $C_2$ for $k = 2$ (blue), $k = 4$ (red), and $k = 6$ (olive) are presented in Fig.~\ref{fig:C2_reheating} for three different reheating temperature values $10^{13}$ GeV (solid), $10^{10}$ GeV (dashed), and $10^7$ GeV (dot-dashed), with $g_* = g_{*s} = 100$, $\phi_{\rm end} = M_{\rm P}$, and $H_{\rm end} = 10^{12}$ GeV. The different lines indicate the upper limit of $|C_2|$. The behaviour here resembles the one we observed for the operator $\mathcal{O}_1$; for the $k = 2$ ($k = 6$) case, the bound becomes weaker (stronger) for lower reheating temperatures, while for the $k = 4$ case, it is independent of the reheating temperature. Unlike in the case of $\mathcal{O}_1$, however, we see from Fig.~\ref{fig:C2_reheating} that the scalar mass range for $|C_2|$ to take the value of $\mathcal{O}(0.01-1)$ is significantly smaller. 
If the scalar mass is in the range of keV--TeV, the reheating temperature is required to be tiny. 

It is also possible to convert the constraints on $C_1$ \eqref{eqn:C1constraint} and $C_2$ \eqref{eqn:C2constraint} into a constraint on the reheating temperature $T_{\rm reh}$, as long as $k \neq 4$ as the reheating temperature dependence drops out for $k=4$, {\it i.e.}, $\Delta_{\rm reh }\rightarrow 1$. Let us, for instance, consider the $k = 2$ case. Recalling that $\Delta_{\rm reh} = \sqrt{H_{\rm end}/H_{\rm reh}}$ and using $H_{\rm reh} = \sqrt{\pi^2 g_* / 90} T_{\rm reh}^2 / M_{\rm P}$, we find from Eq.~\eqref{eqn:C1constraint} that the bound on the reheating temperature associated with the operator $\mathcal{O}_1$ is given by
\begin{align}
	T_{\rm reh}^{(\mathcal{O}_1)} \lesssim
	\frac{1.8 \times 10^{-9}}{C_1^2}
	\left(\frac{g_{*s}}{g_*}\right)
	\left(\frac{{\rm GeV}}{m_s}\right)
	\left(\frac{M_{\rm P}}{H_{\rm end}}\right)
	M_{\rm P}
	\quad \text{for } k = 2
	\,.
\end{align}
Similarly, from Eq.~\eqref{eqn:C2constraint}, we find that the bound on the reheating temperature associated with the operator $\mathcal{O}_2$ is given by
\begin{align}
	T_{\rm reh}^{(\mathcal{O}_2)} \lesssim
	\frac{5.9 \times 10^{-7}}{C_2^2}
	\left(\frac{g_{*s}}{g_*}\right)
	\left(\frac{{\rm GeV}}{m_s}\right)
	\left(
	\frac{H_{\rm end}}{M_{\rm P}}
	\right)^3
	\left(
	\frac{M_{\rm P}}{\phi_{\rm end}}
	\right)^8
	M_{\rm P}
	\quad \text{for } k = 2
	\,.
    \label{eqn:TrehO2k2bound}
\end{align}
We note that the more stringent bound on the reheating temperature comes from the operator $\mathcal{O}_2$.
It is also worth noting that for $k > 4$ cases, we instead have a lower bound on the reheating temperature as the exponent of $\Delta_{\rm reh}$ becomes negative.

\begin{figure}
	\centering
	\includegraphics[width=1\linewidth]{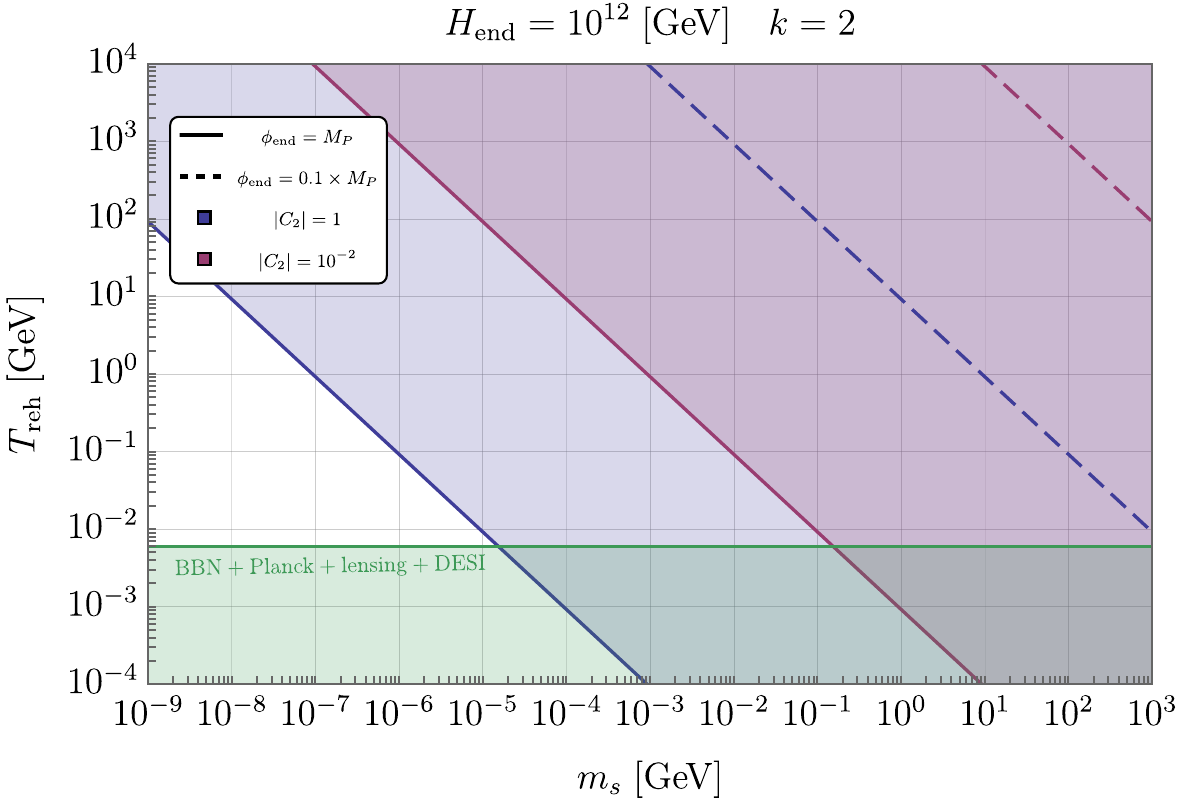}
	\caption{Stringent upper bound on the reheating temperature $T_{\rm reh}$ coming from the operator $\mathcal{O}_2$ in terms of the scalar mass $m_s$ for the quadratic inflaton potential. Two values of the Wilson coefficient, $|C_2| = 1$ (blue) and $|C_2| = 10^{-2}$ (red), and two values of the inflaton field at the end of inflation, $\phi_{\rm end} = M_{\rm P}$ (solid) and $\phi_{\rm end} = 0.1 \times M_{\rm P}$ (dashed), are considered, together with the choice of $g_* = g_{*s} = 100$. The green region indicates the lower bound on the reheating temperature, $T_{\rm reh} \gtrsim 6$ MeV adopted from Ref.~\cite{Barbieri:2025moq}; see also Refs.~\cite{Kawasaki:1999na,Kawasaki:2000en,Hasegawa:2019jsa}.}
	\label{fig:constraintTrehO2k2}
\end{figure}
We present the bound \eqref{eqn:TrehO2k2bound} for the $k = 2$ case in Fig.~\ref{fig:constraintTrehO2k2}. Two values of the Wilson coefficient $C_2$ are considered: $|C_2| = 1$ (blue) and $|C_2| = 10^{-2}$ (red). For the inflaton field value at the end of inflation, $\phi_{\rm end} = M_{\rm P}$ (solid) and $\phi_{\rm end} = 0.1 \times M_{\rm P}$ (dashed) are chosen. One may see that the bound weakens as the couplings and the inflaton field value at the end of inflation become smaller. We remark that for $|C_2| = \mathcal{O}(0.01-1)$ and $\phi_{\rm end} = \mathcal{O}(0.1-1)M_{\rm P}$, the standard scalar freeze-in DM with mass in the keV--TeV range is excluded due to the over-production during reheating. In Fig.~\ref{fig:constraintTrehO2k2}, the green region indicates the lower bound on the reheating temperature, $T_{\rm reh} \gtrsim 6$ MeV \cite{Barbieri:2025moq}; see also Refs.~\cite{Kawasaki:1999na,Kawasaki:2000en,Hasegawa:2019jsa}.

\subsection{Two-stage reheating}
\label{subsec:two-stage-reh-reh}
Having discussed the effects of the quantum gravity-induced, Planck-suppressed operators, $\mathcal{O}_1$ and $\mathcal{O}_2$, on the production of the scalar $s$ during reheating with a generic power-law inflaton potential $V_{\rm inf} \propto \phi^k$, we now examine the extension to a two-stage reheating scenario. The first stage of the post-inflationary phase is considered to be governed by the inflaton potential $V_{\rm inf,1} = \lambda_1 M_{\rm P}^4 (\phi/M_{\rm P})^{k_1}$ right after the end of inflation. After the end of the first stage, the second stage proceeds with the inflaton potential $V_{\rm inf,2} = \lambda_2 M_{\rm P}^4 (\phi/M_{\rm P})^{k_2}$. Denoting the scale factor at the end of the first stage by $a_1$ and the scale factor at the end of the second stage by $a_2$, we note that, by definition, $a_2 > a_1$. Furthermore, the end of the second stage corresponds to the end of reheating, {\it i.e.}, $a_2 = a_{\rm reh}$.

Computing the yield at the end of reheating follows the same procedure as in the single-stage case. In the two-stage reheating case, the power $k$ in the inflaton potential is not a constant but can be regarded as a piecewise function of $a$, changing from $k_1$ to $k_2$. The Boltzmann equation takes the same form as before. Let us first focus on the first operator, $\mathcal{O}_1$, for which the Boltzmann equation is given by Eq.~\eqref{eqn:O1BE-singlestage}.
In the two-stage reheating scenario, we may integrate the Boltzmann equation by splitting the integral into two, obtaining
\begin{align}
	n_s a^3 &\simeq 
	\frac{C_1^2}{8 \pi M_{\rm P}^4} \int_1^a \frac{da'}{a' H} a'^3 \phi_0^4 \omega^4 P
	\nonumber\\ &=
	\frac{\pi C_1^2 M_{\rm P}^4}{32}
	\left(
	f_1
	\int_1^{a_1} da' a'^2
	\frac{\phi_0^{2k_1}}{M_{\rm P}^{2k_1} H}
	+f_2
	\int_{a_1}^a da' a'^2
	\frac{\phi_0^{2k_2}}{M_{\rm P}^{2k_2} H}
	\right)
	\,.
\end{align}
Here, the initial number density $n_s(a_0)$ has again been neglected, we have defined
\begin{align}
	f_i \equiv 
	\left[\frac{\Gamma(1/2+1/k_i)}{\Gamma(1/k_i)}\right]^4
	k_i^4 \lambda_i^2 P_i
	\,,
\end{align}
and $P_1$ and $P_2$ are to be understood as $P \equiv \sum_\ell \ell^4 |(\mathcal{P}^2)_\ell|^2$ for the first and second stages, respectively.
Using the scaling behaviour of $\phi_0$ and $H$, and denoting the value of $\phi_0$ at the end of the first stage by $\phi_1$, we obtain
\begin{align}
	n_s a^3 &\simeq 
	\frac{\pi C_1^2 M_{\rm P}^4}{32}
	\left(
	g_1
	\frac{\phi_{\rm end}^{2k_1}}{M_{\rm P}^{2k_1} H_{\rm end}}
	+g_2
	\frac{\phi_1^{2k_2}}{M_{\rm P}^{2k_2} H_1}
	a_1^3
	\right)
	\,,
\end{align}
where we have taken the $a \gg a_1 \gg 1$ limit and defined
\begin{align}
	g_i \equiv 
	\frac{f_i(k_i+2)}{6(k_i-1)} = 
	\left[\frac{\Gamma(1/2+1/k_i)}{\Gamma(1/k_i)}\right]^4
	\frac{k_i^4(k_i+2)}{6(k_i-1)}
	\lambda_i^2 P_i
	\,.
\end{align}
Recalling $a_1 = (H_{\rm end}/H_1)^{(k_1+2)/(3k_1)}$, together with the definition $\Delta_1 = \sqrt{H_{\rm end}/H_1}$, we find
\begin{align}
	n_s a^3 &\simeq 
	\frac{\pi C_1^2 M_{\rm P}^4}{32}
	\left(
	g_1
	\frac{\phi_{\rm end}^{2k_1}}{M_{\rm P}^{2k_1} H_{\rm end}}
	+g_2
	\frac{\phi_1^{2k_2}}{M_{\rm P}^{2k_2} H_1}
	\Delta_1^{\frac{2(k_1+2)}{k_1}}
	\right)
	\,.
\end{align}
When the inflaton energy density is dominated by the inflaton potential energy at the start and the end of the first stage, the inflaton values, $\phi_{\rm end}$ and $\phi_1$, could be expressed in terms of the Hubble parameters $H_{\rm end}$ and $H_1$. In such a case, the number density of the scalar $s$ can be written as
\begin{align}
	n_s a^3 &\simeq 
	\frac{\pi C_1^2}{32} h_1 H_{\rm end}^3
	\left(
	1
	+\frac{h_2}{h_1}
	\Delta_1^{\frac{4}{k_1}-4}
	\right)
	\,,
\end{align}
where we have defined
\begin{align}
	h_i \equiv \frac{9 g_i}{\lambda_i^2} =
	\left[\frac{\Gamma(1/2+1/k_i)}{\Gamma(1/k_i)}\right]^4
	\frac{3k_i^4(k_i+2)}{2(k_i-1)} P_i
	\,.
\end{align}

The yield at the end of reheating is then given by
\begin{align}
	Y \simeq
	0.043 C_1^2
	\left(\frac{g_*^{3/4}}{g_{*s}}\right)
	h_1
	\left(\frac{H_{\rm end}}{M_{\rm P}}\right)^{\frac{3}{2}}
	\Delta_1^{1-\frac{4}{k_1}}
	\Delta_2^{1-\frac{4}{k_2}}
	\left(
	1
	+\frac{h_2}{h_1}
	\Delta_1^{\frac{4}{k_1}-4}
	\right)
	\,.
\end{align}
One may see that for $H_1 \ll H_{\rm end}$, {\it i.e.}, $\Delta_1 \gg 1$, the production from the first stage dominates as $4/k_1-4 < 0$ under our consideration. We note that the coefficients $h_i$ depend only on $k_i$, and for $k_i$ between 2 and 6, the ratio $h_2/h_1$ takes the value of the order unity, $h_2/h_1 = \mathcal{O}(1)$. It is also worth noting that the production during the second stage depends only on $k_1$. This is because, as we noticed earlier, the production during reheating peaks at the earliest time, being fixed by $a_{\rm end}$ for the first stage and $a_1$ for the second stage. Therefore, the production during the second stage depends only on the duration of the first stage, leading to the independence on $k_2$.
From the abundance constraint \eqref{eqn:abundance-constraint}, we read the following constraint on the coefficient $|C_1|$:
\begin{align}
	|C_1| \lesssim
	1.0 \times 10^{-4}
	\sqrt{\frac{g_{*s}}{g_*^{3/4}}}
	h_1^{-\frac{1}{2}}
	\left(
	\frac{M_{\rm P}}{H_{\rm end}}
	\right)^{\frac{3}{4}}
	\Delta_1^{\frac{4-k_1}{2k_1}}
	\Delta_2^{\frac{4-k_2}{2k_2}}
	\left(
	1
	+\frac{h_2}{h_1}
	\Delta_1^{\frac{4}{k_1}-4}
	\right)^{-\frac{1}{2}}
	\sqrt{\frac{{\rm GeV}}{m_s}}
	\,.
    \label{eqn:C1bound-2stage}
\end{align}

The production of the scalar $s$ through the second operator $\mathcal{O}_2$ in the case of the two-stage reheating can be analysed in the same fashion. For the operator $\mathcal{O}_2$, the Boltzmann equation is given by Eq.~\eqref{eqn:O2BE-singlestage}. Following the same procedure, we find
\begin{align}
	n_s a^3 &\simeq 
	\frac{C_2^2 \phi_{\rm end}^8}{2 \pi M_{\rm P}^4 H_{\rm end}}
	\left[
	\tilde{h}_1
	+\tilde{h}_2\left(
	\frac{\phi_1}{\phi_{\rm end}}
	\right)^8
	\Delta_1^{\frac{4}{k_1}+4}
	\right]
	\,,
\end{align}
where we have taken the $a \gg a_1 \gg 1$ limit and defined
\begin{align}
	\tilde{h}_i \equiv 
	\frac{k_i+2}{6(7-k_i)} Q_i
	\,,
\end{align}
where $Q_1$ and $Q_2$ are to be understood as $Q \equiv \sum_\ell \ell |(\mathcal{P}^4)_\ell|^2$ for the first and second stages, respectively.
We now utilise the continuity of the inflaton potential at the end of the first stage or, equivalently, the start of the second stage, namely at $a = a_1$. Then, $\phi_1/\phi_{\rm end} = (H_1/H_{\rm end})^{2/k_1}$, and thus,
\begin{align}
	n_s a^3 &\simeq 
	\frac{C_2^2 \phi_{\rm end}^8}{2 \pi M_{\rm P}^4 H_{\rm end}} \tilde{h}_1
	\left(
	1
	+\frac{\tilde{h}_2}{\tilde{h}_1}
	\Delta_1^{4-\frac{28}{k_1}}
	\right)
	\,.
\end{align}

The yield at the end of reheating is then given by
\begin{align}
	Y \simeq
	0.069C_2^2
	\left(\frac{g_*^{3/4}}{g_{*s}}\right)
	\tilde{h}_1
	\left(
	\frac{\phi_{\rm end}}{M_{\rm P}}
	\right)^8
	\left(
	\frac{M_{\rm P}}{H_{\rm end}}
	\right)^{\frac{5}{2}}
	\Delta_1^{1-\frac{4}{k_1}}
	\Delta_2^{1-\frac{4}{k_2}}
	\left(
	1
	+\frac{\tilde{h}_2}{\tilde{h}_1}
	\Delta_1^{4-\frac{28}{k_1}}
	\right)
	\,.
    \label{eqn:yield-2stage-O2}
\end{align}
We note that for $k_i$ between 2 and 6, the ratio of the coefficients $\tilde{h}_2/\tilde{h}_1$, which depends only on $k_1$ and $k_2$, varies between 0.1 and 10.
As before, the production from the first stage dominates as long as $H_1$ is sufficiently smaller than $H_{\rm end}$. Moreover, the production during the second stage again depends only on $k_1$, due to the same reason.
From the abundance constraint \eqref{eqn:abundance-constraint}, we read the following constraint on the coefficient $|C_2|$:
\begin{align}
	|C_2| \lesssim
	8.0 \times 10^{-5}
	\sqrt{\frac{g_{*s}}{g_*^{3/4}}}
	\tilde{h}_1^{-\frac{1}{2}}
	\left(
	\frac{M_{\rm P}}{\phi_{\rm end}}
	\right)^{4}
	\left(
	\frac{H_{\rm end}}{M_{\rm P}}
	\right)^{\frac{5}{4}}
	\Delta_1^{\frac{4-k_1}{2k_1}}
	\Delta_2^{\frac{4-k_2}{2k_2}}
	\left(
	1+\frac{\tilde{h}_2}{\tilde{h}_1}
	\Delta_1^{4-\frac{28}{k_1}}
	\right)^{-\frac{1}{2}}
	\sqrt{\frac{{\rm GeV}}{m_s}}
	\,.
    \label{eqn:C2bound-2stage}
\end{align}
\begin{figure}
     \centering
     \includegraphics[width=0.49\linewidth]{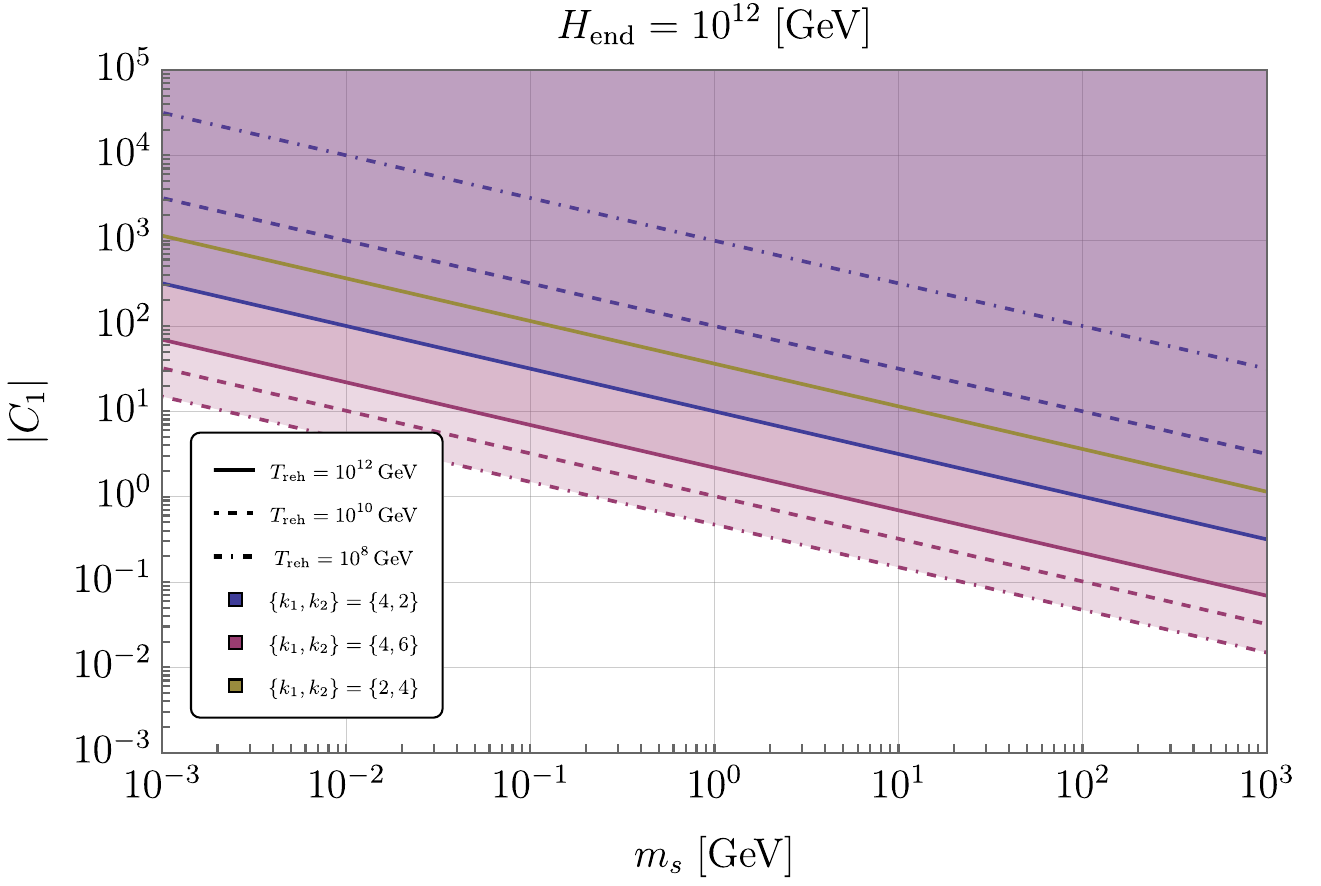}
     \includegraphics[width=0.49\linewidth]{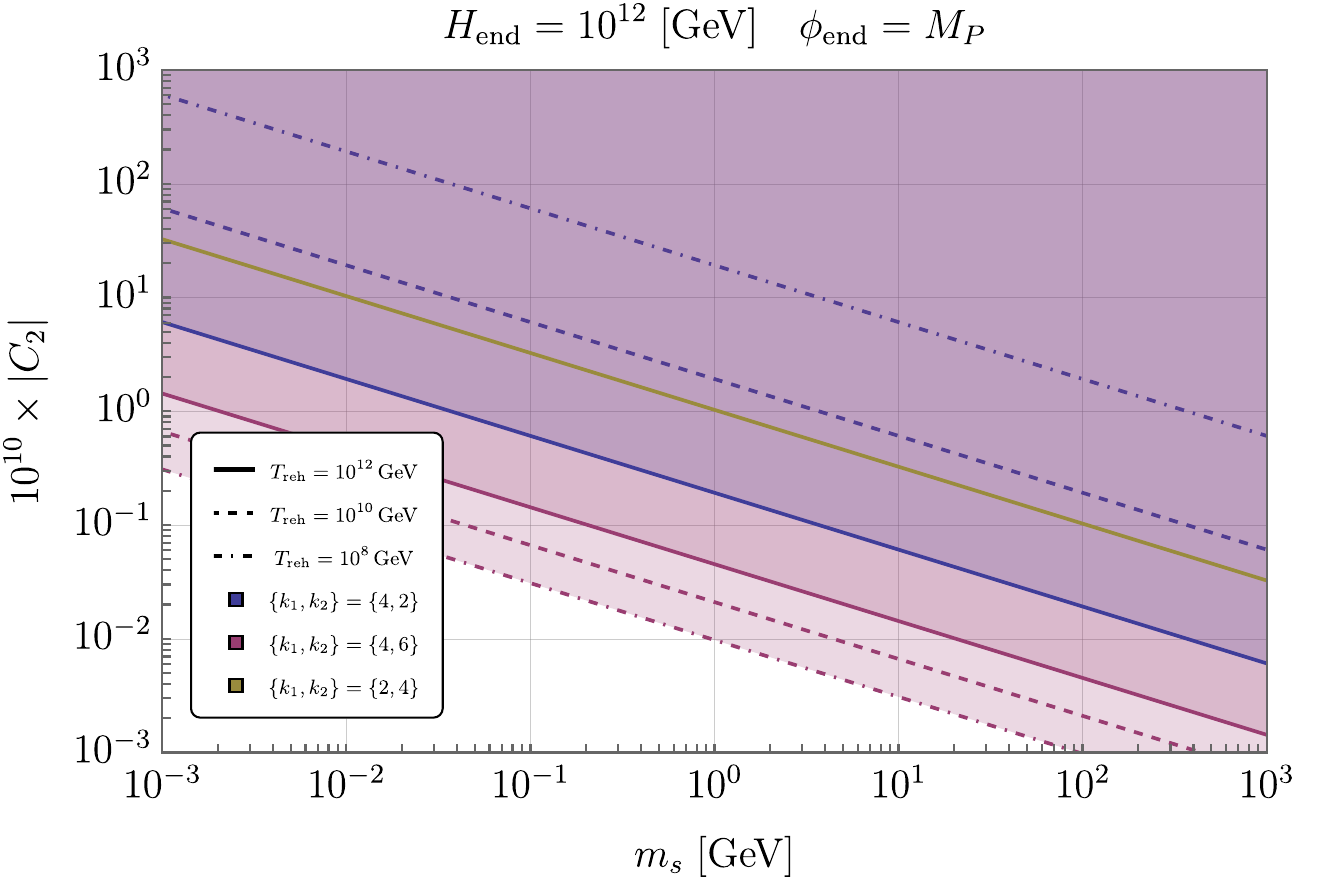}
     \caption{Upper limits on $|C_1|$ (left panel) and $|C_2|$ (right panel) in terms of the scalar mass $m_s$ for two-stage reheating scenarios. The following cases are presented: $\{k_1,k_2\} = \{4,2\}$ (blue), $\{k_1,k_2\} = \{4,6\}$ (red), and $\{k_1,k_2\} = \{2,4\}$ (olive). For each case, three different reheating temperatures are considered, $10^{12}$ GeV (solid), $10^{10}$ GeV (dashed), and $10^8$ GeV (dot-dashed), with $H_1 = 10^8$ GeV, $g_* = g_{*s} = 100$, and $H_{\rm end} = 10^{12}$ GeV.}
     \label{fig:two_stage_prod_C1_C2}
\end{figure}
In Fig.~\ref{fig:two_stage_prod_C1_C2} we present the upper limits on $|C_1|$ \eqref{eqn:C1bound-2stage} (left panel) and $|C_2|$ \eqref{eqn:C2bound-2stage} (right panel) in terms of the scalar mass $m_s$. Considering three different reheating temperatures, namely $10^{12}$ GeV (solid), $10^{10}$ GeV (dashed), and $10^8$ GeV (dot-dashed), the following cases are examined: $\{k_1,k_2\} = \{4,2\}$ (blue), $\{4,6\}$ (red), and $\{2,4\}$ (olive). For the case of $\{k_1,k_2\} = \{4,2\}$, we see that the bound weakens for a longer reheating epoch, which is equivalent to having a longer matter-like stage, aligning with our earlier discussion. The $\{k_1,k_2\} = \{4,6\}$ exhibits an opposite behaviour, also mirroring our previous discussion in Sec.~\ref{subsec:two-stage-reh}.
Finally, the case of $\{k_1,k_2\} = \{2,4\}$ is independent of the reheating temperature. 

\begin{figure}
     \centering
     \includegraphics[width=1\linewidth]{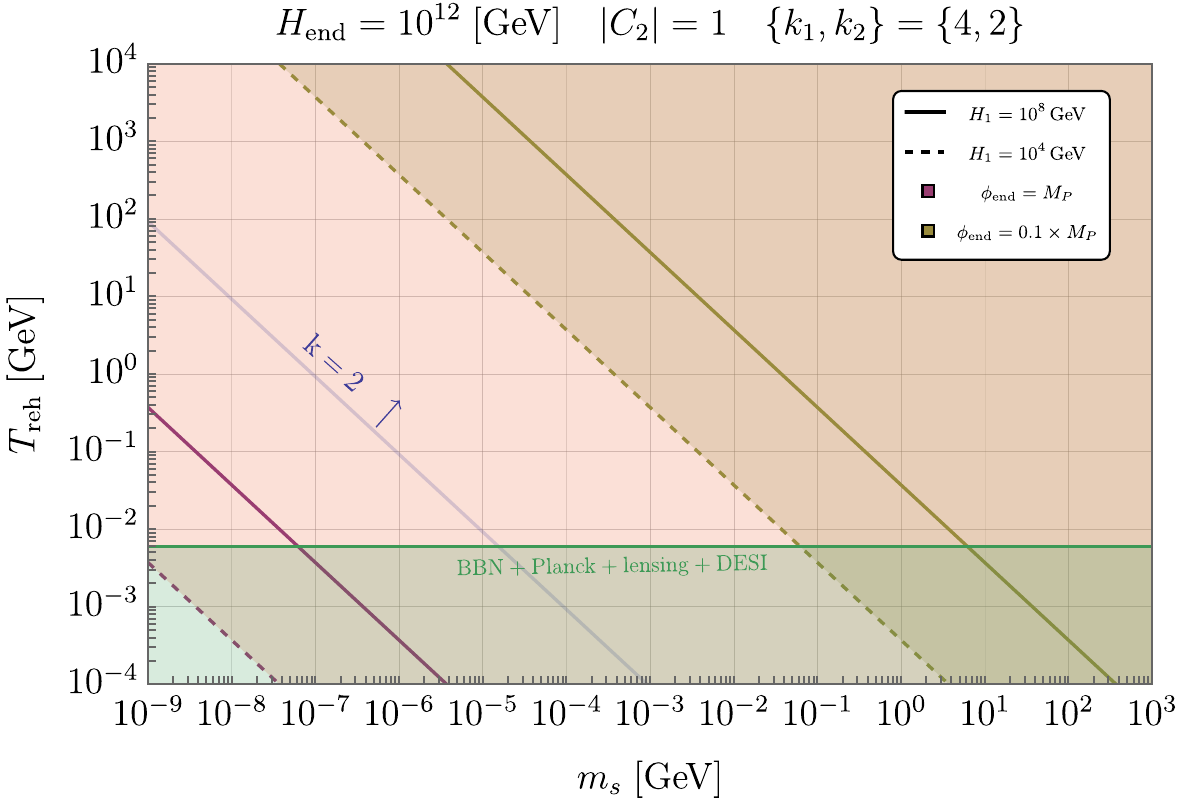}
     \caption{Stringent upper bound on the reheating temperature $T_{\rm reh}$ coming from the operator $\mathcal{O}_2$ in terms of the scalar mass $m_s$ for the $\{k_1, k_2\} = \{4, 2\}$ case. Two values of $H_1$, $10^8$ GeV (solid) and $10^4$ GeV (dashed), and two values of $\phi_{\rm end}$, $M_{\rm P}$ (red) and $0.1M_{\rm P}$ (olive), are considered with $H_{\rm end} = 10^{12}$ GeV, $|C_2| = 1$, and $g_* = g_{*s} = 100$. As a reference, the single-stage reheating scenario with $k = 2$, together with $\phi_{\rm end} = M_{\rm P}$, is indicated with the blue line. We show in green the lower bound on the reheating temperature, $T_{\rm reh} \gtrsim 6$ MeV adopted from Ref.~\cite{Barbieri:2025moq}.}
     \label{fig:two_stage_prod_reh}
\end{figure}
The stringent upper bound on the reheating temperature coming from the operator $\mathcal{O}_2$ is presented in Fig.~\ref{fig:two_stage_prod_reh} in terms of the scalar mass $m_s$ for the $\{k_1, k_2\} = \{4, 2\}$ case. Two values of $H_1$ ($\phi_{\rm end}$) are considered, $H_1 = 10^8$ GeV ($\phi_{\rm end} = M_{\rm P}$) and $H_1 = 10^4$ GeV ($\phi_{\rm end} = 0.1M_{\rm P}$), depicted as solid (red) and dashed (olive) lines. The remaining parameters are fixed as $H_{\rm end} = 10^{12}$ GeV, $|C_2| = 1$, and $g_* = g_{*s} = 100$. As a reference value, the single-stage reheating scenario with $k = 2$, together with $\phi_{\rm end} = M_{\rm P}$, is indicated with the blue line. The green region indicates the lower bound on the reheating temperature, $T_{\rm reh} \gtrsim 6\,\mathrm{MeV}$ from Ref.~\cite{Barbieri:2025moq}.
We observe that for the two-stage reheating scenario with $\{k_1,k_2\}=\{4,2\}$, the bound on the reheating temperature becomes stronger as the value of $H_1$ decreases and the value of $\phi_{\rm end}$ increases. This tendency can be best understood from Eq.~\eqref{eqn:yield-2stage-O2}; given that $\{k_1,k_2\}=\{4,2\}$, the yield is proportional to $T_{\rm reh}$ and $\phi_{\rm end}$, while it is inversely proportional to $H_1$. Increasing $H_1$ thus has the effect of diluting the abundance, while increasing $\phi_{\rm end}$ has the opposite effect, and thus, maintaining the abundance requires a higher reheating temperature.

\subsection{$m$-stage reheating}
\label{subsec:m-stage-reh-reh}
Extending the analysis of the two-stage reheating scenario discussed in Sec.~\ref{subsec:two-stage-reh-reh} to a multi-stage reheating case is straightforward. Considering the scenario where the reheating phase consists of $m$ stages, let us assume the existence of the hierarchy between the Hubble parameters at the end of each stage, that is, $H_{\rm end} \gg H_1 \gg H_2 \gg \cdots \gg H_m = H_{\rm reh}$. In this case, the dominant production comes from the first stage, and the remaining stages act as a dilution or enhancement phase, depending on the value of $k_i$, for the abundance of the produced scalar. 

For the operator $\mathcal{O}_1$, we would then have
\begin{align}
	Y \simeq
	0.043  C_1^2
	\left(\frac{g_*^{3/4}}{g_{*s}}\right)
	h_1
	\left(\frac{H_{\rm end}}{M_{\rm P}}\right)^{\frac{3}{2}}
	\Delta_1^{1-\frac{4}{k_1}}
	\Delta_2^{1-\frac{4}{k_2}}
	\cdots 
	\Delta_m^{1-\frac{4}{k_m}}
	\,,
\end{align}
while for the operator $\mathcal{O}_2$, we get
\begin{align}
	Y \simeq
	0.069 C_2^2
	\left(\frac{g_*^{3/4}}{g_{*s}}\right)
	\tilde{h}_1
	\left(
	\frac{\phi_{\rm end}}{M_{\rm P}}
	\right)^8
	\left(
	\frac{M_{\rm P}}{H_{\rm end}}
	\right)^{\frac{5}{2}}
	\Delta_1^{1-\frac{4}{k_1}}
	\Delta_2^{1-\frac{4}{k_2}}
	\cdots 
	\Delta_m^{1-\frac{4}{k_m}}
	\,.
\end{align}
One may notice that expressions for the yield in the multi-stage reheating scenario could equally be written as
\begin{align}
	Y \simeq Y_{1} 
	\Delta_2^{1-\frac{4}{k_2}}
	\cdots 
	\Delta_m^{1-\frac{4}{k_m}}
	\,,
\end{align}
where $Y_1$ is the yield evaluated at the end of the first stage; see also Eqs.~\eqref{eqn:yield-O1-single} and \eqref{eqn:yield-O2-single}. Indeed, the most efficient production of the scalar $s$ comes from the first stage, and over the course of the remaining stages, the abundance gets diluted if $k_i < 4$ and enhanced if $k_i > 4$. This feature is, in fact, exactly the same as in the production during inflation discussed in Sec.~\ref{sec:prodinf}. The production at an earlier time becomes modulated by subsequent reheating stages.

\section{Discussion and implications}
\label{sec:conc}
In this work, we have investigated the gravitational production of a decoupled scalar $s$ with a generic reheating scenario, encompassing both production during inflation and production during reheating. Building upon previous analyses that focused on simple reheating dynamics, we have systematically examined how a generalised post-inflationary evolution affects the abundance of gravitationally-produced relics, and we have discussed the impact of a low reheating temperature in this context.

For the production during inflation, we have studied five distinct reheating scenarios: instantaneous reheating, matter-dominated reheating with a quadratic inflaton potential, reheating with a generic power-law inflaton potential $V_{\rm inf} \propto \phi^k$, two-stage reheating, and $m$-stage reheating. Our analysis has shown that the behaviour of the constraints coming from the relic abundance strongly depends not only on the power-law index $k$ of the inflaton potential but also on the duration of the reheating stage. For $k < 4$, the reheating phase dilutes the scalar abundance, thereby relaxing the constraints. In this case, prolonging the reheating phase, {\it i.e.}, having a lower reheating temperature, significantly weakens the bounds on the scalar self-interaction coupling $\lambda_s$. For the quartic inflaton potential case of $k = 4$, the abundance becomes independent of the reheating temperature as has been known, resembling the instantaneous reheating scenario. For $k > 4$, corresponding to kination-like eras, the reheating phase enhances the scalar abundance compared to the $k = 4$ case, making the constraints more stringent as the completion of reheating takes longer or, equivalently, the reheating temperature decreases.

The multi-stage reheating scenarios reveal a rich phenomenology. In the two-stage reheating case, for instance, we find that, although the final abundance depends on both the powers of the inflaton potential, $k_1$ and $k_2$, the coefficient and the exponent of $\lambda_s$ in the constraint depend only on the inflaton potential form during the first stage, which is governed by $k_1$. This is because the most efficient production occurs at the earliest times, namely right after the end of inflation. Subsequent reheating stages act as a modulation of this initial production through dilution or enhancement factors. As we have explicitly demonstrated, this feature persists in the general $m$-stage reheating scenario, where the yield can be compactly written as the yield from the first stage multiplied by products of dilution/enhancement factors from each subsequent stage.

For the production during reheating, we have generalised the analysis of quantum gravity-induced operators to generic power-law inflaton potentials and multi-stage scenarios. The two Planck-suppressed operators, $\mathcal{O}_1 = C_1 (\partial s)^2 \phi^2 / M_{\rm P}^2$ and $\mathcal{O}_2 = C_2 \phi^4 s^2 / M_{\rm P}^2$, have been extensively investigated. We have extracted generic constraints on the Wilson coefficients $|C_1|$ and $|C_2|$ which show similar $k$-dependence to the case of production during inflation. Remarkably, for arguably natural values of $|C_1| = \mathcal{O}(0.01-1)$, the GeV--TeV mass range is accessible. Constraints on $|C_2|$ are, however, generally more stringent, particularly for large scalar masses, pointing towards the need of a low reheating temperature for the same mass and coupling values. The production during reheating is UV-dominated as long as $k < 7$, and the major contribution comes from the earliest stages of the post-inflationary evolution. For $k \geq 7$, the production during reheating through operator $\mathcal{O}_2$ becomes IR-dominated rather than UV-dominated, as the number density grows with time. These observations naturally extend to the multi-stage reheating scenarios, where for $k < 7$, the first stage determines the production, and subsequent stages provide dilution/enhancement factors.

Throughout this work, in order to scrutinise the sole effects of gravitational production of a scalar in the early Universe, we have focused on scenarios where the scalar does not directly couple to the SM as well as to the inflaton, thereby gravitational interactions remain to be the only, unavoidable, interaction between the two. However, in realistic models, additional interactions may be present. We briefly comment on these possibilities before we conclude.

The scalar could directly couple to the SM. For instance, one may consider the Higgs-portal coupling of the form $\lambda_{Hs} |H|^2 s^2$, where $H$ is the SM Higgs doublet. The production of a scalar via the Higgs-portal interaction follows the standard freeze-in dynamics, and one typically assumes a negligible initial abundance of the scalar. However, as we discussed in this work, gravitational particle production during reheating may result in over-production, unless the reheating temperature $T_{\rm reh}$ is sufficiently low. In this case, a new regime opens for $m_s > T_{\rm reh}$, where production is Boltzmann-suppressed, allowing substantially larger portal couplings. Controlling the gravitational over-production therefore enables the possibility of observing non-thermally-produced relics in direct, indirect, and collider experiments. Readers may refer to Refs.~\cite{Cosme:2023xpa,Cosme:2024ndc,Costa:2024ugy,Arcadi:2024wwg}, for instance, for more detailed discussion.

One may also consider interactions between the scalar and the inflaton. For instance, a direct coupling of the form $\lambda_{\phi s} \phi^2 s^2$ could be present. Such an inflaton-portal coupling can induce an effective mass of the scalar during inflation. Therefore, the production of the scalar during inflation can be suppressed. The inflaton-portal coupling is, however, known to lead to efficient particle production during reheating; see, {\it e.g.}, Refs.~\cite{Lebedev:2021tas,Choi:2024bdn,Feiteira:2025rpe}. Thus, the issue of over-production remains for the freeze-in mechanism. The interplay between suppression during inflation and enhancement during reheating in the presence of inflaton-portal couplings would require a dedicated analysis.

Finally, a non-minimal coupling to gravity of the form $\xi_s s^2 R$, where $R$ is the Ricci scalar, may modify the production. During the inflationary phase, such a coupling contributes to the effective mass of the scalar $s$, potentially increasing it above the Hubble scale and thereby suppressing quantum fluctuations. Therefore, one may suppress the production during inflation through the non-minimal coupling. However, as shown in Refs.~\cite{Bassett:1997az,Lebedev:2022vwf}, the non-minimal coupling can result in abundant production of the scalar during reheating.

To conclude, our analysis highlights the crucial role of reheating dynamics in assessing the viability of non-thermal DM. We have shown that both inflationary and reheating-era production are highly sensitive to the details of the reheating history. In particular, with low reheating temperatures, the reheating phase could efficiently dilute gravitationally-produced relics for the inflaton potential power-law index less than four, $k < 4$. On the contrary, for $k > 4$, the relic is shown to be enhanced during the reheating phase, putting stringent constraints on the decoupled scalar. In multi-stage reheating, we have shown that the dilution/enhancement effect of subsequent reheating stages factorises. These results emphasise the importance of incorporating not only gravitational effects but also realistic post-inflationary dynamics when confronting feebly-interacting DM scenarios with cosmological constraints and determining the predictivity of non-thermal DM models.

\acknowledgments
We are deeply grateful to O. Lebedev for thoughtful comments on the manuscript.
FC acknowledges partial support from the FORTE project CZ.02.01.01/00/22 008/0004632 co-funded by the EU and the Ministry of Education, Youth and Sports of the Czech Republic. FC acknowledges partial financial support from the Science Foundation Ireland Grant 21/PATHS/9475 (MOREHIGGS) under the SFI-IRC Pathway Programme. JK is supported by National Natural Science Foundation of China (NSFC) under Grant No. 12505079.

\appendix
\section{Non-equilibrium case}
\label{apdx:non-Eq}
We summarise the resultant expressions for the constraint on the scalar self-interaction coupling $\lambda_s$ and the scalar mass $m_s$ coming from the production during inflation when the duration of inflation is not long enough for the probability distribution function of the scalar $s$ to reach its equilibrium.
\begin{itemize}
	\item Instantaneous reheating scenario (see also Ref.~\cite{Lebedev:2022cic}):
	\begin{align}
		m_s \lambda_s^{-\frac{1}{4}}
		\lesssim 
		6.5 \times 10^{-9}
		\left(
		\frac{g_{*s}}{g_{*}^{3/4}}
		\right)
		\left(
		\frac{M_{\rm P}}{H_{\rm end}}
		\right)^{\frac{3}{2}}
		\,{\rm GeV}
	\end{align}

	\item Single-stage reheating scenario with the quadratic inflaton potential (see also Ref.~\cite{Lebedev:2022cic}):
	\begin{align}
		m_s \lambda_s^{-\frac{1}{2}}
		&\lesssim 
		8.6\times 10^{-9}
		\left(
		\frac{g_{*s}}{g_*^{3/4}}
		\right)
		\Delta_{\rm reh}
		\left(
		\frac{M_{\rm P}}{H_{\rm end}}
		\right)^{\frac{3}{2}}
		\, {\rm GeV}
	\end{align}

	\item Single-stage reheating scenario with the quartic inflaton potential (see also Ref.~\cite{Lebedev:2022cic}):
	\begin{align}
		m_s \lambda_s^{-\frac{1}{4}}
		&\lesssim 
		6.5\times 10^{-9}
		\left(
		\frac{g_{*s}}{g_*^{3/4}}
		\right)
		\left(
		\frac{M_{\rm P}}{H_{\rm end}}
		\right)^{\frac{3}{2}}
		\, {\rm GeV}
	\end{align}

	\item Single-stage reheating scenario with the sextic inflaton potential:
	\begin{align}
		m_s \lambda_s^{-\frac{1}{6}}
		&\lesssim 
		5.9\times 10^{-9}
		\left(
		\frac{g_{*s}}{g_*^{3/4}}
		\right)
		\Delta_{\rm reh}^{-\frac{1}{3}}
		\left(
		\frac{M_{\rm P}}{H_{\rm end}}
		\right)^{\frac{3}{2}}
		\, {\rm GeV}
	\end{align}

	\item Two-stage reheating scenario:
	\begin{align}
		m_s \lambda_s^{-\frac{1}{k_1}}
		\lesssim 
		4.9\times 10^{-9} \times 3^{\frac{1}{k_1}}
		\left(
		\frac{g_{*s}}{g_*^{3/4}}
		\right)
		\Delta_1^{\frac{4}{k_1}-1}
		\Delta_2^{\frac{4}{k_2}-1}
		\left(
		\frac{M_{\rm P}}{H_{\rm end}}
		\right)^{\frac{3}{2}}
		\, {\rm GeV}
	\end{align}
\end{itemize}

\section{Computation of $(\mathcal{P}^\gamma)_\ell$}
\label{apdx:compP}
We outline two methods for the computation of $(\mathcal{P}^\gamma)_\ell$. 
Recalling that
\begin{align}
	\phi(t) = \phi_0(t)
	\sum_{\ell=-\infty}^{\infty}
	\mathcal{P}_\ell e^{-i\ell\omega t}
	\,,
	\label{eqn:phi-P}
\end{align}
we have
\begin{align}
	\phi^\gamma &=
	\phi_0^\gamma
	\sum_{\ell_1,\ell_2,\cdots,\ell_\gamma}
	\mathcal{P}_{\ell_1}
	\mathcal{P}_{\ell_2}
	\cdots
	\mathcal{P}_{\ell_\gamma}
	e^{-i(\ell_1+\ell_2+\cdots+\ell_\gamma)\omega t}
	\,.
\end{align}
If we set $\ell=\ell_1+\ell_2+\cdots+\ell_\gamma$, we may re-group the sums that have the same value of $\ell$, obtaining
\begin{align}
	\sum_{\ell_1,\ell_2,\cdots,\ell_\gamma}
	\mathcal{P}_{\ell_1}
	\mathcal{P}_{\ell_2}
	\cdots
	\mathcal{P}_{\ell_\gamma}
	e^{-i(\ell_1+\ell_2+\cdots+\ell_\gamma)\omega t}
	&=
	\sum_{\ell=-\infty}^{\infty}
	\left(
	\sum_{\ell_1+\ell_2+\cdots+\ell_\gamma = \ell}
	\mathcal{P}_{\ell_1}
	\mathcal{P}_{\ell_2}
	\cdots
	\mathcal{P}_{\ell_\gamma}
	\right)
	e^{-i\ell\omega t}
	\,,
\end{align}
where the sum $\sum_{\ell_1+\ell_2+\cdots+\ell_\gamma = \ell}$ represents all combinations of indices that add up to $\ell$. We thus have
\begin{align}
	\phi^\gamma = \phi_0^\gamma
	\sum_{\ell=-\infty}^{\infty}
	\left(
	\mathcal{P}^\gamma
	\right)_\ell e^{-i\ell\omega t}
	\,,
\end{align}
where
\begin{align}
	\left(
	\mathcal{P}^\gamma
	\right)_\ell =
	\sum_{\ell_1+\ell_2+\cdots+\ell_\gamma = \ell}
	\mathcal{P}_{\ell_1}
	\mathcal{P}_{\ell_2}
	\cdots
	\mathcal{P}_{\ell_\gamma}
	\,.
\end{align}
We first present a general method to compute $(\mathcal{P}^\gamma)_\ell$ that works for the generic inflaton potential of the form $V(\phi) \propto \phi^k$. We then provide a potential-dependent method that may be used as a cross-check of the results, focusing on the quadratic ($k = 2$) and quartic ($k = 4$) inflaton potentials.

\subsection{General method}
From Eq.~\eqref{eqn:phi-P}, we see that
\begin{align}
	\frac{\phi(t)}{\phi_0(t)} e^{i \ell' \omega t} = 
	\sum_{\ell=-\infty}^\infty
	\mathcal{P}_\ell 
	e^{-i \ell \omega t} e^{i \ell' \omega t}
	\,.
\end{align}
Integrating it over one full period $T$, we get
\begin{align}
	\int_0^T \frac{\phi(t)}{\phi_0(t)} e^{i \ell' \omega t} dt &= 
	\sum_{\ell=-\infty}^\infty
	\mathcal{P}_\ell
	\int_0^T
	e^{-i \ell \omega t} e^{i \ell'\omega t}
	dt
	\nonumber\\&=
	\sum_{\ell=-\infty}^\infty
	\mathcal{P}_\ell
	T \delta_{\ell\ell'}
	=
	\mathcal{P}_{\ell'} T
	\,.
\end{align}
Here, we have used the fact that $T = 2\pi/\omega$. Therefore, we find
\begin{align}
	\mathcal{P}_\ell =
	\frac{1}{T} \int_0^T
	\frac{\phi(t)}{\phi_0(t)}
	e^{i \ell \omega t} dt
	\,.
\end{align}
We assume that the inflaton oscillates around a symmetric potential. Then, the sine part in $e^{i \ell \omega t}$ goes away, and we are left with
\begin{align}
	\mathcal{P}_\ell =
	\frac{1}{T} \int_0^T
	\frac{\phi(t)}{\phi_0(t)}
	\cos(\ell \omega t) dt
	\,.
\end{align}
As $\phi(t)$ is an even function, we see that $\mathcal{P}_\ell$ becomes zero if $\ell$ is even. Thus, we find
\begin{align}
	\mathcal{P}_\ell =
	\begin{cases}
		0 & \text{ for even }\ell \,,
		\\
		\frac{4}{T} \int_0^{T/4} \frac{\phi(t)}{\phi_0(t)}
		\cos(\ell \omega t) dt &
		\text{ for odd }\ell \,.
	\end{cases}
\end{align}

Let us consider the potential of the form,
\begin{align}
	V(\phi) =
	\lambda M_{\rm P}^4
	\left(\frac{\phi}{M_{\rm P}}\right)^k
	\,,
\end{align}
where $\lambda > 0$ is assumed.
Denoting the value of $\phi$ at the maximum height (or, equivalently, the initial field value) by $\phi_0$, the energy conservation tells us that
\begin{align}
	\frac{1}{2} \dot{\phi}^2 +
	\lambda M_{\rm P}^4
	\left(\frac{\phi}{M_{\rm P}}\right)^k 
	=
	\lambda M_{\rm P}^4
	\left(\frac{\phi_0}{M_{\rm P}}\right)^k
	\,,
\end{align}
where we have neglected the expansion of the Universe whose timescale is much larger than the inflaton oscillation timescale.
Then, we obtain
\begin{align}
	\dot{\phi}^2 =
	\frac{2\lambda}{M_{\rm P}^{k-4}}
	\left(
	\phi_0^k - \phi^k
	\right)
	\,.
\end{align}
Assuming that the inflaton is rolling down the potential from $\phi>0$, we may re-write this equation as
\begin{align}
	\dot{\phi} =
	-\sqrt{\frac{2\lambda}{M_{\rm P}^{k-4}}}
	\sqrt{\phi_0^k - \phi^k}
	\,.
\end{align}
Integrating it then gives
\begin{align}
	t(\phi) =
	\sqrt{\frac{M_{\rm P}^{k-4}}{2\lambda}}
	\int_{\phi}^{\phi_0} \frac{d\phi'}{\sqrt{\phi_0^k - \phi'^k}}
	=
	\sqrt{\frac{M_{\rm P}^{k-4}}{2\lambda}}\frac{1}{\phi_0^{k/2-1}}
	\int_x^1 \frac{dx'}{\sqrt{1-x'^k}}
	\,,
\end{align}
where we have set $x \equiv \phi/\phi_0$.
We then note that
\begin{align}
	\mathcal{P}_{\ell,{\rm odd}} = 
	\sqrt{
		\frac{M_{\rm P}^{k-4}}{2\lambda}
	}
	\frac{1}{\phi_0^{k/2-1}}
	\frac{2\omega}{\pi}
	\int_0^1 \frac{x}{\sqrt{1 - x^k}}
	\cos\left(
	\frac{\ell \omega}{\phi_0^{k/2-1}}
	\sqrt{\frac{M_{\rm P}^{k-4}}{2\lambda}}
	\int_{x}^1 \frac{dx'}{\sqrt{1-x'^k}}
	\right)
	dx
	\,.
\end{align}
Since $m_\phi^2 = V''(\phi=\phi_0) = k(k-1)\lambda M_{\rm P}^{4-k} \phi_0^{k-2}$, and thus,
\begin{align}
	\phi_0^{\frac{k}{2}-1} = \frac{m_\phi}{\sqrt{k(k-1)\lambda M_{\rm P}^{4-k}}}
	\,,
\end{align}
we may write
\begin{align}
	\mathcal{P}_{\ell,{\rm odd}} &= 
	\frac{\sqrt{2k(k-1)}}{\pi}
	\frac{\omega}{m_\phi}
	\int_0^1 \frac{x}{\sqrt{1 - x^k}}
	\cos\left[
	\ell \sqrt{\frac{k(k-1)}{2}} \frac{\omega}{m_\phi}
	\int_{x}^1 \frac{dx'}{\sqrt{1-x'^k}}
	\right]
	dx
	\nonumber\\&=
	\frac{k}{\sqrt{\pi}}
	\frac{\Gamma(1/2+1/k)}{\Gamma(1/k)}
	\int_0^1 \frac{x}{\sqrt{1 - x^k}}
	\cos\left[
	\frac{\ell k \sqrt{\pi}}{2}
	\frac{\Gamma(1/2+1/k)}{\Gamma(1/k)}
	\int_{x}^1 \frac{dx'}{\sqrt{1-x'^k}}
	\right]
	dx
	\,,
\end{align}
where we have used
\begin{align}
	\omega = 
	m_\phi 
	\sqrt{\frac{\pi k}{2(k-1)}}
	\frac{\Gamma(1/2+1/k)}{\Gamma(1/k)}
	\,.
\end{align}
Note that the integral inside the cosine function is the hypergeometric function, ${}_2 F_1$. Thus, we arrive at
\begin{align}
	\mathcal{P}_{\ell,{\rm odd}} &= 
	\frac{k}{\sqrt{\pi}}
	\frac{\Gamma(1/2+1/k)}{\Gamma(1/k)}
	\nonumber\\&\quad\times
	\int_0^1 \frac{x}{\sqrt{1 - x^k}}
	\cos\left[
	\frac{\ell k \sqrt{\pi}}{2}
	\frac{\Gamma(1/2+1/k)}{\Gamma(1/k)}
	\left(
	\frac{\sqrt{\pi}\Gamma(1+1/k)}{\Gamma(1/2+1/k)}
	-x {}_2 F_1\left(
	\frac{1}{2}, \frac{1}{k}; 1+\frac{1}{k};
	x^k
	\right)
	\right)
	\right]
	dx
	\,.
\end{align}
Restricting to the integer-$k$ case simplifies the expression as follows:
\begin{align}
	\mathcal{P}_{\ell,{\rm odd}} &= 
	\frac{k}{\sqrt{\pi}}
	\frac{\Gamma(1/2+1/k)}{\Gamma(1/k)}
	\nonumber\\&\quad\times
	\int_0^1 \frac{x}{\sqrt{1 - x^k}}
	\cos\left[
	\frac{\ell k \sqrt{\pi}}{2}
	\frac{\Gamma(1/2+1/k)}{\Gamma(1/k)}
	\left(
	\frac{\sqrt{\pi}\Gamma(1+1/k)}{\Gamma(1/2+1/k)}
	-\frac{1}{k}\,\mathrm{B}\left(x^k;\frac{1}{k},\frac{1}{2}\right)
	\right)
	\right]
	dx
	\,.
\end{align}
where $B$ is the incomplete Beta function.

\subsection{Potential-dependent method}
Let us now compute $(\mathcal{P}^\gamma)_\ell$ in a potential-dependent fashion. We start from the quadratic inflaton potential, $k = 2$. In the case of the quadratic potential, we have $\phi = \phi_0 \cos(m_\phi t)$, where $m_\phi$ is the mass of $\phi$. Thus, we simply have $\mathcal{P} = \cos(m_\phi t)$. Let us evaluate $(\mathcal{P}^2)_\ell$. From
\begin{align}
	\mathcal{P}^2 = \cos^2(m_\phi t) = 
	\frac{1}{2} + \frac{1}{4} e^{2 i m_\phi t} + \frac{1}{4} e^{-2 i m_\phi t}
	\,,
\end{align}
we read
\begin{align}
	(\mathcal{P}^2)_0 = \frac{1}{2}
	\,,\quad
	(\mathcal{P}^2)_{\pm 2} = \frac{1}{4}
	\,.
\end{align}
Similarly, we can evaluate $(\mathcal{P}^4)_\ell$. From
\begin{align}
	\mathcal{P}^4 = \cos^4(m_\phi t) = 
	\frac{3}{8}
	+\frac{1}{4} e^{2 i m_\phi t}
	+\frac{1}{4} e^{-2 i m_\phi t}
	+\frac{1}{16} e^{4 i m_\phi t}
	+\frac{1}{16} e^{-4 i m_\phi t}
	\,,
\end{align}
we find
\begin{align}
	(\mathcal{P}^4)_0 = \frac{3}{8}
	\,,\quad
	(\mathcal{P}^4)_{\pm 2} = \frac{1}{4}
	\,,\quad
	(\mathcal{P}^4)_{\pm 4} = \frac{1}{16}
	\,.
\end{align}

Next, we consider the quartic potential,
\begin{align}
	V(\phi) = \lambda \phi^4
	\,.
\end{align}
From the energy conservation, we read
\begin{align}
	\frac{1}{2} \dot{\phi}^2 + \lambda \phi^4 =
	\lambda \phi_0^4
	\,,
\end{align}
where $\phi_0$ denotes the value of $\phi$ at the maximum height, or, equivalently, the initial field value, and we have neglected the expansion of the Universe as stated earlier. We then find
\begin{align}
	\dot{\phi}^2 =
	2 \lambda \left(
	\phi_0^4 - \phi^4
	\right)
	\,.
\end{align}
Assuming that the field is rolling down the potential from $\phi>0$, we may re-write this equation as
\begin{align}
	\dot{\phi} =
	-\sqrt{2\lambda}
	\sqrt{\phi_0^4 - \phi^4}
	\,.
\end{align}
Integrating it leads to
\begin{align}
	t(\phi) =
	\sqrt{2\lambda}
	\int_{\phi}^{\phi_0} \frac{d\phi'}{\sqrt{\phi_0^4 - \phi'^4}}
	=
	\frac{1}{\sqrt{2\lambda}\phi_0}
	\int_x^1 \frac{dx'}{\sqrt{1-x'^4}}
	\,,
\end{align}
where $x \equiv \phi/\phi_0$. The integral gives the elliptic function of the first kind,
\begin{align}
	\int \frac{dx}{\sqrt{1-x^4}} = F(\sin^{-1}(x)| -1)\,.
\end{align}
Therefore,
\begin{align}
	t = 
	\frac{1}{\sqrt{2\lambda}}
	\frac{1}{\phi_0}
	\left[
	F(\pi/2|-1) - F(\sin^{-1}(x)|-1)
	\right]
	\,,
\end{align}
or, equivalently,
\begin{align}
	F(\sin^{-1}(x)|-1) =
	F(\pi/2|-1) - \sqrt{2\lambda}\phi_0 t
	\,.
\end{align}
Inverting it, we obtain
\begin{align}
	\phi = \phi_0 sn\left(
	F(\pi/2|-1)-\sqrt{2\lambda}\phi_0 t
	\bigg\vert-1\right)
	\,,
\end{align}
where $sn$ is the Jacobi elliptic function $sn$.
The Jacobi elliptic function $sn(u|v)$ is periodic with period $4K(v)$, where $K(v)$ is the complete elliptic integral of the first kind. The $sn(u|v)$ function can be series-expanded as \cite{DLMF:Jacobi}
\begin{align}
	sn(u|v) = \frac{2\pi}{\sqrt{v}K(v)}
	\sum_{n=0}^{\infty}
	\frac{q^{n+1/2}}{1-q^{2n+1}}
	\sin\left[
	(2n+1)\frac{\pi u}{2K(v)}
	\right]
	\,,
\end{align}
where
\begin{align}
	q = \exp\left[
	-\pi \frac{K(1-v)}{K(v)}
	\right]
	\,.
\end{align}
We note that
\begin{align}
	F(\pi/2|-1) = K(-1)
	\,,
\end{align}
and
\begin{align}
	K(-1) = \frac{\sqrt{\pi}\Gamma(1/4)}{4\Gamma(3/4)}
	\,.
\end{align}
Now, using
\begin{align}
	\sin(x) = \frac{1}{2i}\left(
	e^{ix} - e^{-ix}
	\right)\,,
\end{align}
we find
\begin{align}
	sn(u|v) = \frac{\pi}{\sqrt{v}K(v)}
	\sum_{n=0}^{\infty}
	\frac{q^{n+1/2}}{i(1-q^{2n+1})}
	\left[
	\exp\left(i \frac{(2n+1) \pi u}{2K(v)}\right)
	-\exp\left(-i \frac{(2n+1) \pi u}{2K(v)}\right)
	\right]
	\,.
\end{align}
If we set $\ell = 2n+1$, then we can write
\begin{align}
	sn(u|v) = \frac{\pi}{\sqrt{v}K(v)}
	\sum_{\ell={\rm odd}}
	\frac{q^{\ell/2}}{i(1-q^\ell)}
	\left[
	\exp\left(i \frac{\ell \pi u}{2K(v)}\right)
	-\exp\left(-i \frac{\ell \pi u}{2K(v)}\right)
	\right]
	\,.
\end{align}
Thus, we obtain
\begin{align}
	\phi &= \phi_0 sn\left(
	F(\pi/2|-1)-\sqrt{2\lambda}\phi_0 t
	\bigg\vert-1\right)
	=
	\phi_0 sn\left(
	K(-1)-\sqrt{2\lambda}\phi_0 t
	\bigg\vert-1\right)
	\nonumber\\&=
	4 \phi_0 \sqrt{\pi} \frac{\Gamma(3/4)}{\Gamma(1/4)}
	\sum_{\ell={\rm odd}}
	\frac{q^{\ell/2}}{(1-q^\ell)}
	\exp\left(-i \frac{\ell \pi}{2}\right)
	\exp\left(i \ell \sqrt{8\pi \lambda} \frac{\Gamma(3/4)}{\Gamma(1/4)} \phi_0 t \right)
	\nonumber\\&\quad
	-4 \phi_0 \sqrt{\pi} \frac{\Gamma(3/4)}{\Gamma(1/4)}
	\sum_{\ell={\rm odd}}
	\frac{q^{\ell/2}}{(1-q^\ell)}
	\exp\left(i \frac{\ell \pi}{2} \right)
	\exp\left(-i \ell \sqrt{8 \pi \lambda} \frac{\Gamma(3/4)}{\Gamma(1/4)} \phi_0 t \right)
	\,.
\end{align}
Here,
\begin{align}
	q = \exp\left[
	-\pi \frac{K(2)}{K(-1)}
	\right]
	\,.
\end{align}
Recalling that the oscillation frequency is given by
\begin{align}
	\omega = \sqrt{8\pi \lambda} \phi_0 \frac{\Gamma(3/4)}{\Gamma(1/4)}
	\,,
\end{align}
we find
\begin{align}
	\phi = \phi_0
	\sum_{\ell=-\infty}^{\infty}
	\mathcal{P}_\ell e^{-i\ell\omega t}
	\,,
\end{align}
where
\begin{align}
	\mathcal{P}_{\ell>0,{\rm odd}} &=
	4 \sqrt{\pi} \frac{\Gamma(3/4)}{\Gamma(1/4)}
	\frac{q^{\ell/2}}{(1-q^\ell)}
	\exp\left(-i \frac{\ell \pi}{2}\right)
	\,,\\
	\mathcal{P}_{\ell<0,{\rm odd}} &= 
	-4 \sqrt{\pi} \frac{\Gamma(3/4)}{\Gamma(1/4)}
	\frac{q^{\ell/2}}{(1-q^\ell)}
	\exp\left(i \frac{\ell \pi}{2}\right)
	\,,\\
	\mathcal{P}_{\ell,{\rm even}} &= 
	0
	\,.
\end{align}
One may further simplify these expressions. Noting that $K(2)/K(-1) = 1 - i$, we read $q = -e^{-\pi}$. Then,
\begin{align}
	\frac{q^{\ell/2}}{(1-q^\ell)}
	\exp\left(-i \frac{\ell \pi}{2}\right)
	=
	\frac{e^{-\ell\pi/2}}{1-(-1)^\ell e^{-\ell\pi}}
	\,.
\end{align}
Similarly, we find
\begin{align}
	\frac{q^{\ell/2}}{(1-q^\ell)}
	\exp\left(i \frac{\ell \pi}{2}\right) =
	\frac{(-1)^\ell e^{-\ell\pi/2}}{1-(-1)^\ell e^{-\ell\pi}}
	\,.
\end{align}
Therefore, we conclude that for the quartic inflaton potential case,
\begin{align}
	\mathcal{P}_{\ell,{\rm odd}} &=
	4 \sqrt{\pi} \frac{\Gamma(3/4)}{\Gamma(1/4)}
	\frac{e^{-\ell\pi/2}}{1-(-1)^\ell e^{-\ell\pi}}
	\,,\\
	\mathcal{P}_{\ell,{\rm even}} &= 
	0
	\,.
\end{align}
We note that
\begin{align}
	\mathcal{P}_{\pm 1} \approx 0.4775
	\,,\quad
	\mathcal{P}_{\pm 3} \approx 0.0215
	\,,\quad
	\mathcal{P}_{\pm 5} \approx 0.0009
	\,,\quad
	\cdots
\end{align}

For $(\mathcal{P}^2)_\ell$, we get
\begin{align}
	\left(\mathcal{P}^2\right)_0 \approx 0.457
	\,,\quad
	\left(\mathcal{P}^2\right)_{\pm 2} \approx
	0.2486
	\,,\quad
	\left(\mathcal{P}^2\right)_{\pm 4} \approx 0.0214
	\,,\quad\cdots
\end{align}
and thus, $P \equiv \sum_{\ell=1}^\infty \ell^4 |(\mathcal{P}^2)_\ell|^2 \approx 1.11$. For $(\mathcal{P}^4)_\ell$, we have
\begin{align}
	\left(\mathcal{P}^4\right)_0 \approx 
	0.3333
	\,,\quad
	\left(\mathcal{P}^4\right)_{\pm 2} \approx
	0.2379
	\,,\quad
	\left(\mathcal{P}^4\right)_{\pm 4} \approx 
	0.0821
	\,,\quad\cdots
\end{align}
and thus, $Q \equiv \sum_{\ell=1}^\infty \ell |(\mathcal{P}^4)_\ell|^2 \approx 0.141$.


\bibliographystyle{JHEP}
\bibliography{main}
\end{document}